\documentclass[11pt,a4paper]{article}
\usepackage{jheppub, braket, tikz, dsfont}

\makeatletter
\newcommand\para{\@startsection{paragraph}{4}{\z@}{-3.25ex plus
-1ex minus -.2ex}{1.5ex plus .2ex}{\normalsize\itshape}}
\makeatother

\def\K{{\cal K}}
\def\affK{\widehat{\cal K}}

\def\L{{\cal L}}

\def\V{{\cal V}}

\def\1{{\bf 1}}
\def\Z{{\mathbb{Z}}}

\def\psl{\mathfrak{psl}(2|2)}
\def\affpsl{\widehat{\mathfrak{psl}}(2|2)_k}
\def\g{\mathfrak{g}}
\def\affg{\widehat{\mathfrak{g}}}
\def\bg{\mathfrak{g}^{(0)}}

\def\res{\mathrm{Res}}
\def\ch{\mathrm{ch}}

\newcommand{\mycaption}[1]{\caption{\textsl{#1}}}

\newcommand\no[1]{\left\langle#1\right\rangle}

\title{String States on $\rm \bf AdS_3 \times S^3$ from the Supergroup}

\author{Sebastian Gerigk}

\affiliation{
Institut f\"ur Theoretische Physik, ETH Zurich, \\
CH-8093 Z\"urich, Switzerland}

\emailAdd{gerigk@itp.phys.ethz.ch} 

\abstract{In the hybrid formulation of string theory on $\rm{AdS}_3 \times \rm{S}^3$, the compactification-independent physical spectrum at the first massive level is determined. We find that these states transform in representations of the Lie superalgebra $\psl$ and present a cohomological characterisation of them within the context of the $\rm PSL(2|2)$ WZW model. We conjecture that  a similar cohomological description of the physical states also exists at higher mass levels and give partial evidence for this by comparing our results to the RNS string spectrum.}

\begin{document}

\maketitle

\section{Introduction}

Due to its conjectured correspondence to conformal field theories \cite{Aharony:1999ti,Maldacena:1997re,Witten:1998qj}, superstring theory in Anti-de-Sitter gravitational backgrounds has attracted a lot of interest lately. A particular accessible background is $\rm AdS_3 \times S^3$ with pure NSNS-flux \cite{Maldacena:1998bw}. In that case, the world sheet theory possesses an RNS world sheet description in terms of a WZW model \cite{deBoer:1998fk,Evans:1998qu,Giveon:1998ns,Kutasov:1999xu} and hence tools of conformal field theory can be used to check the correspondence to a large extent \cite{Cardona:2009hk,Gaberdiel:2007vu,Giribet:2007wp,Pakman:2007hn,Taylor:2007hs}. However, the inclusion of RR-flux to the string background is problematic from the world sheet perspective and it has been mostly discussed at the classical level \cite{Park:1998un,Pesando:1998wm,Rahmfeld:1998zn} (see also \cite{Yu:1998qw} for a treatment at the quantum level in light cone gauge).

The hybrid formulation \cite{Berkovits:1999im} aims to resolve this problem\footnote{We should mention that there also is a complementary approach to understanding string theory on $\rm AdS_3$ including RR-flux by using its integrability \cite{Babichenko:2009dk}. We thank the referee for pointing out this reference to us.} as it is a covariantly quantised description of string theory with manifest target space supersymmetry in six dimensions, which are eventually taken to describe $\rm AdS_3 \times S^3$. In fact, it has been argued that string theory on $\rm AdS_3 \times S^3$ is appropriately described by a nonlinear $\sigma$-model on the supergroup $\rm PSL(2|2)$. These models are conformally invariant even without the inclusion of the Wess-Zumino term \cite{Bershadsky:1999hk}, and hence one obtains a two-dimensional moduli space of conformal field theories. From the point of view of string theory on $\rm AdS_3 \times S^3$, these moduli are associated with possible flux backgrounds including both RR- and NSNS-flux. In particular, the pure NSNS-flux background is associated with the WZW point in moduli space \cite{Berkovits:1999im}. The WZW model on $\rm PSL(2|2)$ has been studied extensively \cite{Gotz:2005ka,Gotz:2006qp} and it has been found that the underlying conformal field theory is logarithmic, as it  is quite generally the case for WZW~models with supergroup target spaces \cite{Hikida:2007sz,Saleur:2006tf,Schomerus:2005bf}.

It has been known for some time that the hybrid formulation of string theory correctly reproduces the RNS spectrum of massless physical states \cite{Dolan:1999dc,Dolan:2000yp,Langham:2000ac}, \textit{i.e.}\ the supergravity spectrum on $\rm AdS_3 \times S^3$ \cite{deBoer:1998ip,Deger:1998nm}. Recently, this analysis has been extended to take into account the logarithmic nature of the conformal field theory underlying the $\rm PSL(2|2)$ WZW model \cite{Gaberdiel:2011vf,Troost:2011fd}. It is the goal of this paper to
investigate the connection between string theory on $\rm AdS_3 \times S^3$ and the $\rm PSL(2|2)$~WZW~model further by identifying the correct spectrum of massive compactification-independent physical string states within the $\rm PSL(2|2)$ WZW model. The strategy is similar to the one followed in the massless case: the physical state constraints of the hybrid formulation are applied to vertex operators that correspond to states which carry one affine excitation of either the WZW currents or the additional hybrid bosons $\rho$, $\sigma$ and $H$. The latter boson $H$ is associated with the ${\rm U}(1)$ current of the ${\cal N}=2$ superconformal algebra on the world sheet in the RNS formulation. We find that the physical state constraints eliminate any state that carries affine excitations of these hybrid bosons. Thus the physical state constraints reduce to conditions on the WZW spectrum only. Furthermore, physical string states are found to be characterised by a intriguingly simple procedure: at the first affine level of the WZW spectrum, one has to first identify the subspace of Virasoro primaries and then take the cohomology with respect to an appropriately chosen normal ordered product of WZW currents. This description can be naturally extended to any mass level and we propose that this characterisation indeed determines the complete physical string spectrum. The proposal is confirmed at the first two mass levels by comparing the resulting spectrum to the spectrum obtained in the RNS framework. 

The paper is organised as follows. We start by introducing some notation and review the representation theory of $\psl$ as well as the basics of the hybrid formulation in section~\ref{review}. The physical state constraints of the hybrid formulation are applied to appropriately chosen vertex operators in section~\ref{comp_ind_constr}, where the algebraic properties of the relevant normal ordered fields in the $\rm PSL(2|2)$ WZW model are also discussed. The resulting algebraic characterisation of physical string states is then applied to the representations of $\psl$ and the spectrum is shown to be in agreement with the RNS spectrum in section \ref{corr_rns}. In section~\ref{conjecture} we finally give our proposal how physical string states could be characterised at all mass levels and confirm it at the second level. Section~\ref{conclusion} contains our conclusions. Our conventions as well as further technical details are summarised in four appendices.

\section{Review of Basic Concepts and Notation} \label{review}

Before going in medias res, let us set the stage by introducing some notation. Suppose $\phi$ and $\psi$ are elements in some space of states. We write the vertex operator associated to the state $\psi$ as $\psi(z) \equiv V(\psi,z)$ and the modes as $\psi_n \equiv V_n(\psi)$. The latter are defined by a Laurent expansion of the form 
\begin{equation}
V(\psi,z) = \sum_n V_n(\psi) z^{-n-h_\psi} \,.
\end{equation}
In terms of vertex operators the operator product expansion takes the form (see \textit{e.g.}\ \cite{Goddard:1989dp, Gaberdiel:1999mc}),
\begin{equation}
V(\phi, z) V(\psi, w) = \sum_l V(V_{-l-h_\phi}(\phi) \psi,w) (z-w)^l \,.
\end{equation}
Given an OPE $\phi(z)\psi(w)$, we shall denote the poles by curly brackets \cite{DiFrancesco:1997nk},
\begin{equation}
\phi(z)\psi(w) = \sum_{l} \{\phi_{-l-h_\phi} \psi\}(w)(z-w)^l \,,
\end{equation}
so the curly brackets are just a shorthand notation for $ \{\phi_{-l-h_\phi} \psi\}(w) = V(V_{-l-h_\phi}(\phi) \psi,w)$. The \textit{normal ordered product} is simply the nonsingular part of the OPE, denoted by angle brackets,
\begin{equation}
\no{\phi(z)\psi(w)} \equiv \phi(z)\psi(w) - \text{poles in $(z-w)$} \,.
\end{equation}
We further write $\no{\phi\psi}(w) \equiv \lim_{z \rightarrow w} \no{\phi(z)\psi(w)}$. The normal ordered product has in general to be distinguished from the creation-annihilation ordering of the modes that will be denoted by colons, $:\hspace{-1mm}\phi\psi\hspace{-1mm}:(w)$. Finally, we will mostly be interested in the first order pole of an OPE, which coincides with the residue of that OPE at the singular locus. Hence we denote the first order pole of $\phi(z) \psi(w)$ by $\res\, \phi(z) \psi(w)$. 

\subsection{The Lie superalgebra $\mathfrak{psl}(2|2)$ and Kac modules} \label{intro_psl}

Let us begin by reviewing the representation theory of $\g = \psl$ and its affine version $\widehat{\g}$; this will also allow us to fix our notations.\footnote{In the succeeding we follow closely \cite{Gaberdiel:2011vf}.}

Like any Lie superalgebra, $\psl$ allows for a decomposition into bosonic and fermionic generators 
$\g = \bg \oplus \g^{(1)}$ \cite{Scheunert:1979xy}, where $\bg$ is the bosonic Lie subalgebra 
$\bg = \mathfrak{sl}(2) \oplus \mathfrak{sl}(2)$. Furthermore, $\psl$ is a Lie superalgebra of type I, 
which means that the fermionic summand $\g^{(1)}$ can be further decomposed as 
$\g^{(1)} = \g_{-1} \oplus \g_{1}$ such that
\begin{equation}
\{\g_{-1},\g_1\} \subset \bg \ , \qquad \{\g_1,\g_1\} = \{\g_{-1},\g_{-1}\} = 0 \ .
\end{equation}
This decomposition introduces a natural grading $\varrho$, where $\varrho(\g_{\pm1}) = \pm 1$ and 
$\varrho(\bg) = 0$. $\varrho$ lifts to a $\Z$-grading on the universal enveloping algebra ${\cal U}(\g)$ in the 
obvious way. An explicit description of the generators and their commutation relations can be extracted from 
Appendix~\ref{app:A}.

For comparison to string theory on AdS$_3 \times {\rm S}^3$ we will mainly be interested in 
representations whose decomposition which respect to the bosonic subalgebra 
$\bg = \mathfrak{sl}(2) \oplus \mathfrak{sl}(2)$ leads to infinite-dimensional discrete series 
representations with respect to the first $\mathfrak{sl}(2)$ (that describes isometries
on ${\rm AdS}_3$),  and finite-dimensional representations with respect to the second 
$\mathfrak{sl}(2)$ (that describes isometries on ${\rm S}^3$).
As in \cite{Gotz:2006qp} we shall label them by a doublet of half-integers 
$\lambda = (j_1, j_2)$ where $j_1 \leq - \tfrac{1}{2}$ and $j_2 \geq 0$. The cyclic state of the corresponding
representation is then characterised by 
\begin{equation}
\begin{array}{c}
J^0 \ket{\lambda} = j_1 \ket{\lambda} \ , \qquad K^0 \ket{\lambda} = j_2 \ket{\lambda} \ , \\[0.25cm]
J^+ \ket{\lambda} = K^+ \ket{\lambda}=\left(K^-\right)^{(2j_2 + 1)}  \ket{\lambda} = 0 \ .
\end{array}
\end{equation}
Here $J^0,J^\pm$ are the generators of the first $\mathfrak{sl}(2)$ with commutation relations 
\begin{equation}
{}[J^0,J^\pm] = \pm J^\pm \ , \qquad [J^+,J^-] = 2 J^0 \ ,
\end{equation}
while $K^0,K^\pm$ are the generators of the second $\mathfrak{sl}(2)$ that satisfy identical commutation 
relations. We denote the corresponding highest weight representation of 
$\bg = \mathfrak{sl}(2) \oplus \mathfrak{sl}(2)$ by $\V(\lambda)$. 

Each representation $\V(\lambda)$ of $\bg$ gives rise to a  representation of the full Lie superalgebra
by taking all the modes in $\g_{+1}$ to act trivially on all states in $\V(\lambda)$, $\g_{+1} \V(\lambda)=0$, and 
by taking the modes in $\g_{-1}$ to be the fermionic creation operators. The resulting representation is usually 
called the Kac module  \cite{Kac:1977hp} and will be denoted by $\K(\lambda)$. The Kac module can be in turn
decomposed in representations of the bosonic subalgebra, which is illustrated in Fig. \ref{fig:Kac}. There we 
defined the weights
\begin{equation}
\lambda^\alpha_\beta = (j_1 + \tfrac{\alpha}{2}, j_2 + \tfrac{\beta}{2}) \,,
\end{equation}
so \textit{e.g.} $\lambda^+_- = (j_1 + \frac{1}{2},j_2-\frac{1}{2})$ and $\lambda^{++} = (j_1+1, j_2)$. In fact, this set of weights will be encountered several times throughout this work.

The grading $\varrho$ of the universal enveloping algebra ${\cal U}(\g)$ induces a grading on the Kac module, 
where we take all states in $\V(\lambda)$ to have the same grade, say $g\in \Z$.  If we want to stress this 
grade assignment, we shall sometimes write $\K_g(\lambda)$. The states involving one fermionic generator 
from $\g_{-1}$ applied to the states in $\V(\lambda)$  then have grade $g-1$, {\it etc}.

\begin{figure}[htb] 
\begin{center}
\begin{tikzpicture}[scale=0.8]


\node (K1_u) at (12,12) {${\cal V}(\lambda)_0$};
\node (K1_l) at (10,10) {${\cal V}(\lambda^+_+)_{-1}$};
\node (K1_r) at (14,10) {${\cal V}(\lambda^-_-)_{-1}$};
\node (K1_d) at (12,8) {$2\, {\cal V}(\lambda)_{-2}$};

\draw (K1_u) to (K1_l);
\draw (K1_u) to (K1_r);
\draw (K1_d) to (K1_l);
\draw (K1_d) to (K1_r);


\node (K2_u) at (6,10) {${\cal V}(\lambda^+_-)_{-1}$};
\node (K2_l) at (4,8) {${\cal V}(\lambda_{--})_{-2}$};
\node (K2_r) at (8,8) {${\cal V}(\lambda^{++})_{-2}$};
\node (K2_d) at (6,6) {${\cal V}(\lambda^+_-)_{-3}$};

\draw (K2_u) to (K2_l);
\draw (K2_u) to (K2_r);
\draw (K2_d) to (K2_l);
\draw (K2_d) to (K2_r);


\node (K3_u) at (18,10) {${\cal V}(\lambda^-_+)_{-1}$};
\node (K3_r) at (20,8) {${\cal V}(\lambda^{--})_{-2}$};
\node (K3_l) at (16,8) {${\cal V}(\lambda_{++})_{-2}$};
\node (K3_d) at (18,6) {${\cal V}(\lambda^-_+)_{-3}$};f

\draw (K3_u) to (K3_l);
\draw (K3_u) to (K3_r);
\draw (K3_d) to (K3_l);
\draw (K3_d) to (K3_r);


\node (K4_l) at (10,6) {${\cal V}(\lambda^+_+)_{-3}$};
\node (K4_r) at (14,6) {${\cal V}(\lambda^-_-)_{-3}$};
\node (K4_d) at (12,4) {${\cal V}(\lambda)_{-4}$};

\draw (K1_d) to (K4_l);
\draw (K1_d) to (K4_r);
\draw (K4_d) to (K4_l);
\draw (K4_d) to (K4_r);


\draw (K1_u) to (K2_u);
\draw (K1_l) to (K2_r);
\draw (K1_r) to (K3_r);
\draw (K1_d) to (K2_d);

\draw (K1_u) to (K3_u);
\draw (K1_l) to (K3_l);
\draw (K1_r) to (K2_l);
\draw (K1_d) to (K3_d);

\draw (K1_d) to (K2_u);
\draw (K4_l) to (K2_r);
\draw (K4_r) to (K2_l);
\draw (K4_d) to (K2_d);

\draw (K1_d) to (K3_u);
\draw (K4_l) to (K3_l);
\draw (K4_r) to (K3_r);
\draw (K4_d) to (K3_d);

\end{tikzpicture}
\end{center}

\mycaption{Decomposition of the Kac-module ${\cal K}(\lambda)$ into $\g^{(0)}$-representations. The lines indicate the action of the fermionic generators $\g^{(1)}$.} \label{fig:Kac} 
\end{figure}

The characters of Kac modules have been discussed in \cite{Gotz:2006qp}. However, in this work, we will be only interested in the highest weight states with respect to the bosonic subalgebra $\mathfrak{g}^{(0)} \simeq \mathfrak{sl}(2) \oplus \mathfrak{sl}(2)$. Let us define the branching function\footnote{Technically, one would rather refer to the function $K_{\lambda}(x,y)$ as the generating function of branching rules for the decomposition of $\K(\lambda)$ into $\g^{(0)}$-representations.} of a Kac module ${\cal K}(\lambda)$ as
\begin{equation}
K_{\lambda}(x,y) = \mathrm{Tr}^{(0)}_{\K(\lambda)} \left( x^{J^0} y^{K^0} \right) \,,
\end{equation}
where the trace $\mathrm{Tr}^{(0)}$ is taken over highest weight states with respect to $\mathfrak{g}^{(0)}$ only. Using Fig. \ref{fig:Kac}, it is straightforward to evaluate the trace and we obtain
\begin{align}
K_{\lambda}(x,y) &= x^{j_1} y^{j_2}\Bigl(2\,x^{\frac{1}{2}} y^{\frac{1}{2}}+2\, x^{\frac{1}{2}}y^{-\frac{1}{2}}+2\,x^{-\frac{1}{2}} y^{\frac{1}{2}}+2\,x^{-\frac{1}{2}} y^{-\frac{1}{2}}+x+x^{-1}+y+y^{-1}+4 \Bigr) \nonumber \\ 
 &= x^{j_1} y^{j_2} \left(x^{\frac{1}{2}} + y^{\frac{1}{2}} + x^{-\frac{1}{2}} + y^{-\frac{1}{2}}\right)^2 \,.
\end{align}
Note that 
\begin{equation}
x^{l_1} y^{l_2} K_{\lambda}(x,y) = K_{\lambda'}(x,y) \qquad \text{with} \quad \lambda' = (j_1 + l_1, j_2 + l_2) \,.
\end{equation}

Whenever the quadratic Casimir $C_2(\lambda)$ vanishes, \textit{i.e.}\ when $\lambda = (-j-1,j)$ or $\lambda = (j,j)$ (see appendix \ref{app:A}), the Kac module $\K(\lambda)$ is reducible and called atypical. In that case, one can take an appropriate quotient in order to arrive at an irreducible representation. Since these kinds of representations will not be of major importance in the present work, we refer to \cite{Gaberdiel:2011vf} for a detailed discussion. However, let us just state that the adjoint representation of $\g$ is such an atypical, but finite-dimensional representation; we will denoted it by $\L_1(\frac{1}{2})$ since its cyclic state has weight $\lambda = (\frac{1}{2},\frac{1}{2})$ and grading $+1$.

The affine version of $\psl$ at affine level $k$ is denoted by $\widehat{\mathfrak{g}} = \affpsl$ and its commutation relations are listed in Appendix \ref{app:A}. The zero modes of $\widehat{\mathfrak{g}}$ define a $\psl$ subalgebra within $\widehat{\mathfrak{g}}$, commonly referred to as the horizontal subalgebra. In abuse of notation, the horizontal subalgebra will also be called $\mathfrak{g}$. It should be clear from the context whether it refers to the Lie superalgebra $\psl$ or the horizontal subalgebra of $\affpsl$.

Affine Lie superalgebras allow for a decomposition into eigenspaces with respect to the adjoint action of the Virasoro zero-mode $L_0$ which is obtained from the Sugawara construction. It is the so-called level decomposition, which should not to be confused with the affine level $k$ of the affine Lie superalgebra $\affg$,
\begin{equation}
\widehat{\mathfrak{g}}\quad \simeq\quad \bigoplus_{n \in \mathbb{Z}} \,\widehat{\mathfrak{g}}_n \,,
\end{equation}
where $\widehat{\mathfrak{g}}_n$ is the vector space spanned by all modes of mode number $n$. Each direct summand transforms in the adjoint representation $\L_1(\frac{1}{2})$ under the adjoint action of the horizontal subalgebra $\affg_0 = \mathfrak{g}$.

Representations of $\affg$ are most easily generated in a similar way as Kac modules are generated in the case of Lie superalgebras. We start with a set of states that transform as a Kac module ${\cal K}(\lambda)$ under the action of the horizontal subalgebra $\g$. These states are called the affine ground states and ${\cal K}(\lambda)$ is the affine ground state representation. In order to expand the set of affine ground states to a representation of the full affine algebra $\affg$, we let all positive modes annihilate the states in ${\cal K}(\lambda)$, $\affg_{n}\, {\cal K}(\lambda) = 0$ for $n \geq 1$. The negative modes act freely on the affine ground states modulo commutation relations. The resulting representation is called an affine Kac module of weight $\lambda$, $\widehat{\K}(\lambda)$. It contains additional singular vectors if the weight $\lambda = (j_1, j_2)$ satisfies $j_1-j_2 \in k\, \Z$ or $j_1 + j_2 + 1 \in k\, \Z$ \cite{Gotz:2005ka}. As we have explained, for applications to string theory, we are interested in the case where $j_1 \leq -\frac{1}{2}$ and $j_2 \geq 0$. Furthermore, consistency of string theory requires the spin values to be bounded by $-\frac{k}{2}-1 < j_1$ \cite{Evans:1998qu} and $j_2 \leq \frac{k}{2}$. So for $k \geq 3$, the only remaining condition for the appearance of singular vectors are $j_1+j_2+1=0$, which means that already the ground state representation contains singular vectors. It has been argued that in this case all affine submodules are generated from singular vectors in the affine ground state representation \cite{Gotz:2005ka}. Since for massive string states we will see that $j_1 + j_2 + 1 \neq 0$, all affine Kac modules that we will encounter in the following are irreducible.

Like the affine Lie superalgebra $\affg$ itself, affine Kac modules $\affK(\lambda)$ allow for a level decomposition as well,
\begin{equation}
\affK(\lambda) \quad \simeq \quad \bigoplus_{n \in \mathbb{N}}\, \affK^{(n)}(\lambda) \,,
\end{equation}
where $\affK^{(n)}(\lambda)$ is the $L_0$-eigenspace of eigenvalue $\frac{1}{2k} C_2(\lambda) + n$. Note that $\frac{1}{2k} C_2(\lambda)$ is the \mbox{$L_0$-eigenvalue} of the ground state representation. Because of $[L_0, \g] = 0$, each direct summand yields a representation of the horizontal subalgebra $\g$. Clearly, $\affK^{(0)}(\lambda) \simeq \K(\lambda)$ as $\g$-representations. Since $\affg_1$ transforms in the adjoint representation, the first level of the Kac-modules decomposes under the action of $\g$ as
\begin{align}
\affK^{(1)}_0(\lambda)\Bigr|_\g \quad &\simeq \quad \L_1(\tfrac{1}{2}) \otimes \K_0(\lambda) \nonumber \\
&\simeq \quad 2\, {\cal K}_0(\lambda) \oplus {\cal K}_0(\lambda^{++}) \oplus {\cal K}_0(\lambda^{--}) \oplus {\cal K}_0(\lambda_{++}) \oplus {\cal K}_0(\lambda_{--}) \nonumber \\[0.25cm]
&\hspace{3cm}\oplus \,  \bigoplus_{\alpha,\beta = \pm} \left({\cal K}_{-1}(\lambda^\alpha_\beta) \oplus {\cal K}_{+1}(\lambda^\alpha_\beta)\right) \,. \label{first_affine_lvl} 
\end{align}
Here we have added subscripts to the modules keeping track of the grading for later use. The above result on the tensor product of a finite-dimensional with an infinite-dimensional representations of $\g$ agrees with the expectation one might have gained from the analysis of tensor products of finite-dimensional representations \cite{Gotz:2005ka}. Further evidence for the decomposition in (\ref{first_affine_lvl}) can be obtained by considering the characters of the representations on both sides of the equation. Let $\ch_{\L(\frac{1}{2})}(x,y)$ and $\ch_{\K(\lambda)}(x,y)$ denote the characters of $\L(\frac{1}{2})$ and $\K(\lambda)$, respectively:
\begin{align}
\ch_{\L(\frac{1}{2})}(x,y) &= \sum_{(\mu_1, \mu_2) \in \Lambda\left(\L(\frac{1}{2})\right)} x^{\mu_1} y^{\mu_2} \,,\\
 \ch_{\K(\lambda)}(x,y) &= \underbrace{\frac{x^{j_1}}{1-x^{-1}} \frac{\left(y^{-j_2}-y^{j_2+1}\right)}{1-y}}_{= \ch_{\V(\lambda)}(x,y)} \left(x^{\frac{1}{2}}+ y^{\frac{1}{2}}+x^{-\frac{1}{2}}+ y^{-\frac{1}{2}} \right)^2 \,,
\end{align}
where  $\Lambda\left(\L(\frac{1}{2})\right)$ denotes the set of weights of $\L(\frac{1}{2})$ including multiplicities, \textit{i.e.} the eigenvalues of the $\psl$ generators in appendix \ref{app:A} under the adjoint action of $J^0$ and $K^0$. These characters satisfy the relation
\begin{equation}
\ch_{\L(\frac{1}{2})}  \ch_{\K(\lambda)} = \sum_{\mu \in \Lambda\left(\L(\frac{1}{2})\right)} \ch_{\K(\lambda + \mu)} \,.
\end{equation}
which implies a decomposition of the form (\ref{first_affine_lvl}) assuming that $\L(\frac{1}{2}) \otimes \K(\lambda)$ is fully reducible. This will be the case if the quadratic Casimir $C_2(\lambda)$ is a non-zero integer, since then $j_1$ is generically not a half-integer, and therefore $j_1 + j_2 + n \neq 0$ for all $n \in \Z$. Then all weights appearing in the decomposition (\ref{first_affine_lvl}) are typical, which in turn implies full reducibility. We also constructed the cyclic states of any Kac module in the direct summand explicitly.\footnote{In order to find the cyclic states in $\affK^{(1)}(\lambda)$, we have decomposed $\affK^{(1)}(\lambda)$ into weight spaces with respect to the zero modes $J^0_0$ and $K^0_0$. In contrast to the full space $\affK^{(1)}(\lambda)$, these weight spaces are finite-dimensional and hence it is possible to write down an explicit basis for each of them. It is then just a matter of linear algebra to evaluate the cyclic state conditions $J^+_0 \psi = K^+_0 \psi = S^{\alpha\beta}_{+,0} \psi = 0$ on each of these weight spaces.}

The $\g$-representations appearing at the second level can be found by noting that these states are generated from the affine ground states by either acting with bilinears of elements in $\affg_1$ or a single element of $\affg_2$, where the latter is again transforming in the adjoint representation $\L(\frac{1}{2})$ of $\g$. Bilinears of elements in $\affg_1$ transform as the symmetric part\footnote{Since the representations contain fermionic as well as bosonic states, the symmetric part of the tensor product should be understood as the antisymmetric combination whenever both entries are fermionic.} of the tensor product representation $\L(\frac{1}{2}) \otimes \L(\frac{1}{2})$, which is \cite{Gotz:2005ka}
\begin{equation}
\mathrm{Sym}\Bigl( \L(\tfrac{1}{2}) \otimes \L(\tfrac{1}{2}) \Bigr) \quad \simeq \quad \K(0,1) \oplus \K(1,0) \oplus {\bf 1} \,,
\end{equation}
where ${\bf 1}$ denotes the trivial representation of $\g$, associated to the trace. Hence the decomposition of the states at the second level into $\g$-representations is given by
\begin{equation}
\affK^{(2)}(\lambda)\Bigr|_{\g} \simeq \bigg[\Bigl({\cal K}(0,1) \oplus {\cal K}(1,0)\Bigr) \otimes {\cal K}(\lambda)\biggr] \, \oplus {\cal K}(\lambda) \, \oplus \biggl[ \,{\cal L}(\tfrac{1}{2}) \otimes {\cal K}(\lambda) \biggr]\,. \label{second_affine_lvl}
\end{equation}
The tensor products on the right hand side can, of course, be evaluated if necessary. For example, the first direct summand decomposes as
\begin{align}
\Bigl({\cal K}(0,1) \oplus {\cal K}(1,0)\Bigr) \otimes {\cal K}(\lambda) \simeq & \phantom{2\, \oplus} 12\, \K(\lambda) \oplus  {\cal K}(\lambda^{4+}) \oplus {\cal K}(\lambda^{4-}) \oplus {\cal K}(\lambda_{4+}) \oplus {\cal K}(\lambda_{4-}) \nonumber \\[0.25cm]
&\oplus\ 6\, \Bigl({\cal K}(\lambda^{++}) \oplus {\cal K}(\lambda^{--}) \oplus {\cal K}(\lambda_{++}) \oplus {\cal K}(\lambda_{--})\Bigr)  \nonumber \\[0.25cm]
&\oplus\ 2\left( \bigoplus_{|\alpha| + |\beta| = 4 \atop |\alpha|,|\beta| \geq 1} {\cal K}(\lambda^\alpha_\beta)\right) \oplus\ 8 \left(  \bigoplus_{\alpha,\beta = \pm} {\cal K}(\lambda^\alpha_\beta) \right) \,.
\end{align}
The decomposition of the remaining tensor product in (\ref{second_affine_lvl}) has already been given in (\ref{first_affine_lvl}). In a similar fashion, the decomposition at higher level can be determined.

\subsection{The hybrid formulation of strings on  AdS$_3\times {\rm S}^3$}

In \cite{Berkovits:1999im}, the worldsheet fields of the covariantly gauge fixed RNS-string were successfully redefined such that manifest target space supersymmetry in six dimensions is obtained. In particular, it has been shown that RNS string theory on AdS$_3\times {\rm S}^3 \times M$ is equivalent to a $\rm PSL(2|2)$ WZW model plus a topologically twisted superconformal field theory describing the internal manifold $M$ equipped with a topologically twisted ${\cal N}=2$ superconformal structure on the world sheet. This formulation of string theory is now commonly referred to as the hybrid formulation of string theory or simply the hybrid string. Following \cite{Berkovits:1999im}, we denote the ${\cal N}=2$ superconformal generators of the compactification CFT on $M$ by $\{T_C, G^\pm_C, J_C\}$. Then the generators of the full ${\cal N}=2$ superconformal algebra are defined by
\begin{eqnarray}
 T &= &T^\mathrm{WZW}  -\tfrac{1}{2}\left((\partial\rho)^2 + (\partial\sigma)^2\right) + \tfrac{3}{2} \partial^2(\rho + i \sigma) + T_C\,, \label{SCA_N=2_1} \\
 G^+ &=& -e^{-2\rho-i\sigma}\, P + e^{-\rho} Q + e^{i\sigma}{\cal T} + {G^+_C} \\
 G^- &=& e^{-i\sigma} + {G^-_C} \,, \\
 J &=&  \partial(\rho + i\sigma) + J_C \,, \label{SCA_N=2_4}
\end{eqnarray}
where the fields $\rho$ and $\sigma$ are free bosons often referred to as $\rho\sigma$-ghosts,
\begin{equation}
\rho(z) \rho(w) \sim - \ln(z-w) \,, \qquad \sigma(z) \sigma(w) \sim -\ln (z-w) \,,
\end{equation}
and $T^\mathrm{WZW}$ is the Sugawara energy momentum tensor of the $\rm PSL(2|2)$ WZW model. The fields $P$, $Q$ are normal ordered products of the $\rm PSL(2|2)$ WZW currents\footnote{The field $Q$ as defined in (\ref{def_Q}) differs slightly from the expression in \cite{Berkovits:1999im} since it has a different numerical factor in front of the second term. However, only if $Q$ takes the form (\ref{def_Q}), one obtains a physical string spectrum that fits the RNS spectrum on the massive level. Since $\no{S^a_- \partial S^a_-}$ vanishes on affine ground states, this makes no difference at the massless level.},
\begin{align}
P &= \tfrac{1}{24} \varepsilon_{abcd} \no{S^a_- S^b_-S^c_-S^d_-} = \no{S^1_- S^2_-S^3_-S^4_-} \,, \\
Q &= \frac{1}{2\sqrt{k}} \left[ \no{K_{ab} \no{S^a_- S^b_-}} + 4 i \no{S^a_- \partial S^a_-} \right] \,. \label{def_Q}
\end{align}
Here and in the remainder of this work, summation over $\mathfrak{so}(4)$-indices $a,b,c,\ldots$ that appear twice is understood, independent of whether they are upper or lower indices. Note that in general the normal ordered product in the $\rm PSL(2|2)$ WZW model is neither commutative nor associative, so the sequence of normal ordered products in the trilinear term of $Q$ is crucial. However, the sequence is not important in $P$ as fermionic currents of the same grading anticommute. Therefore, the sequence is not specified in $P$ in order not to clutter notation. Finally, the field ${\cal T}$ in $G^+$ is the Sugawara energy momentum tensor $T^\mathrm{WZW}$ deformed by the $U(1)$-current $\partial(\rho+i\sigma)$,
\begin{align}
{\cal T} &= T^\mathrm{WZW} -\tfrac{1}{2}(\partial \rho + i \partial \sigma)^2 + \tfrac{1}{2} \partial^2(\rho + i \sigma) \,.
\end{align}
The exponentials in the $\rho\sigma$-ghost fields, say $e^{m\rho+in\sigma}$, are easily checked to have $U(1)$-charge $n-m$ and conformal weight $\frac{1}{2}(m-n)(3-m-n)$. Using that $P$ and $Q$ are fields of conformal weight $4$ and $3$, respectively, this implies that $G^+$ and $G^-$ have conformal weight $1$ and $2$, as expected in a topologically twisted ${\cal N}=2$ superconformal field theory. 

The zero-mode of $Q$ is special in the sense that it has particularly interesting commutation relations with the generators of $\g$. It is clear that $Q_0$ commutes with the bosonic subalgebra since all $\mathfrak{so}(4)$-indices are contracted. Furthermore, it is not difficult to convince oneself that $Q_0$ also commutes with the generators of negative grading $\g_{-1}$. However, $Q_0$ does not commute with $\g_{+1}$. Fortunately, it turns out that the commutator of $Q_0$ with elements in $\g_{+1}$ takes a particularly nice form,
\begin{equation}
\left[Q_0, S^{\alpha\beta}_{+,0}\right] =-i\sqrt{k} \no{S^{\alpha\beta} T^\mathrm{WZW}}_0 = -i\sqrt{k} \sum_{n \in \mathbb{Z}} :S^{\alpha\beta}_{-,-n}L_{n}:  \,, \label{comm_rel_QS}
\end{equation}
where the $L_n$ correspond the modes of the energy momentum tensor $T^\mathrm{WZW}$. The important point to extract form this commutation relation is that, when we are considering a $\g$-module and restrict ourselves to the submodule of Virasoro primary states of vanishing conformal weight, $Q_0$ commutes with $\g_{+1}$ modulo Virasoro descendants. Assuming that the $\g$-module decomposes into a direct sum of Virasoro primaries and Virasoro descendants, this implies that $Q_0$ projected onto the submodule of Virasoro primaries induces a $\g$-homomorphism on that submodule. We will see that due to this property of $Q_0$, the algebraic structure of $\psl$ is still intact on shell, at least for the compactification-independent spectrum. In particular, physical states transform in representations of $\g$ and hence algebraic tools can be used to describe them. For details on the deduction of (\ref{comm_rel_QS}) as well as an explicit realisation of $Q_n$ in terms of $\affpsl$-modes, we refer the interested reader to appendix \ref{app_Q}.

It is possible to lift the ${\cal N}=2$ superconformal algebra in (\ref{SCA_N=2_1}) - (\ref{SCA_N=2_4}) to a small ${\cal N}=4$ superconformal algebra \cite{Berkovits:1994vy}. The additional generators are
\begin{align}
 \tilde{G}^- &= -e^{-3\rho-2i\sigma-iH}\, P + e^{-2\rho-i\sigma-iH} Q + e^{-\rho-iH}{\cal T} + {\tilde{G}^-_C} \\
 \tilde{G}^+ &= e^{\rho+iH} + {\tilde{G}^+_C} \,, \\
 J^{++} &= e^{\rho+i\sigma+iH} \,, \\
 J^{--} &= e^{-\rho-i\sigma-iH} \,,  
\end{align}
where the free boson $H$ is defined by $J_C = i \partial H$ and hence satisfies the OPE
\begin{equation}
H(z)H(w) \sim -2\,\ln(z-w) \,.
\end{equation}
The ${\cal N}=4$ algebra plays an important role for defining physical string states in the context of the hybrid formulation. Indeed, it turns out that with respect to the extended ${\cal N}=4$ superconformal algebra, physical string states coincide with so-called ${\cal N}=4$ topological string states \cite{Berkovits:1999im} defined by
\begin{equation}
G^+_0 \phi^+ = \tilde{G}^+_0 \phi^+ = (J_0 - 1) \phi^+ = L_0 \phi^+ = 0 \,, \qquad \phi^+ \sim \phi^+ + G^+_0 \tilde{G}^+_0 \Lambda^- \,.
\end{equation}
The superscript on the states indicates the $U(1)$-charge of the respective state.

\section{Compactification-Independent Hybrid String States at First Level} \label{comp_ind_constr}

In this section, we will analyse the lightest massive string states in the hybrid formulation that are independent of the choice of compactification manifold $M$. The results we present here extend previous findings on the description of the massless compactification-independent string states. It was argued in \cite{Berkovits:1999im, Dolan:1999dc} that these former states are given by elements in the cohomology of the zero-mode $Q_0$ on affine ground states of conformal weight zero. By using representation theory of the Lie superalgebra $\mathfrak{psl}(2|2)$, this result was worked out in the context of the $\rm PSL(2|2)$ WZW model from an algebraic point of view in \cite{Gaberdiel:2011vf,Troost:2011fd}, and the resulting spectrum was shown to match the supergravity answer \cite{deBoer:1998ip} exactly.

Similar to the massless case, our goal in this paper is to find a description that allows us to identify physical compactification-independent massive string states within the $\rm PSL(2|2)$ WZW model as well. This would imply that it is not necessary to work with the complicated ${\cal N}=2$ superconformal structure (\ref{SCA_N=2_1}) - (\ref{SCA_N=2_4}), which plays an important role in the hybrid formulation, in order to determine the physical string spectrum. Rather we only have to understand the algebraic structure of the affine Lie superalgebra $\widehat{\mathfrak{psl}}(2|2)$ and its representations. 

Our strategy to achieve this is to start in the hybrid formulation and to note that hybrid vertex operators factorise into a vertex operator of the $\rm PSL(2|2)$ WZW model, and a vertex operator containing the ghost fields $\rho$, $\sigma$ as well as fields of the ${\cal N}=4$ superconformal algebra on the compactification manifold. Evaluating the physical state conditions on these vertex operators result in conditions on the $\rm PSL(2|2)$ WZW vertex operator alone, thus reducing the hybrid description to an algebraic description in the context of the $\rm PSL(2|2)$ WZW model.

\subsection{Evaluation of the Hybrid Physical States Constraints} \label{ev_hybrid_constraints}

Our goal is to find an appropriate description of the massive string states within the $\rm PSL(2|2)$ WZW model. From the hybrid formulation we know that physical states have vanishing conformal weight, so the quadratic Casimir $C_2(\lambda)$ of the horizontal subalgebra of $\mathfrak{psl}(2|2)$ has to be negative on the ground state representation. We will be considering here states at the first level, whose ground states therefore satisfy 
\begin{equation}
\frac{1}{2k}C_2(\lambda) = L_0 = -1 \,. \label{lambda_condition}
\end{equation}
Hence any affine descendant has conformal weight greater or equal to zero. Note that (\ref{lambda_condition}) defines a set of allowed weights for affine Kac-modules. 

Recall that in the massless case \cite{Berkovits:1999im, Dolan:1999dc}, the vertex operators of compactification-independent physical states were of the form
\begin{equation}
V^+_0 \equiv \no{\phi\, e^{2 \rho + i \sigma} J^{++}_C} = \no{\phi\, e^{2 \rho + i \sigma + iH}} \,, \label{vo_massless}
\end{equation}
where $\phi$ is a vertex operator of the target space supersymmetric theory in six dimensions, \textit{i.e.} of the $\rm PSL(2|2)$ WZW model. The guiding principle in arguing for this ansatz is that it must have vanishing conformal weight\footnote{Recall that the ${\cal N}=4$ SCFT with target space $M$ is topologically twisted. Thus $J^{++}_C$ has conformal weight zero. One might wonder why not to consider an ansatz without any excitations on $M$, $\no{\phi\, e^{2 \rho + i \sigma}}$. However, a analysis along the same lines as presented here leads to the conclusion that such a vertex operator would not contain any physical degrees of freedom.}, unit $U(1)$ charge and it must be massless from the point of view of the six-dimensional theory, \textit{i.e.} $L^\mathrm{WZW}_0 \phi = 0$. The reader might wonder why $V^+_0$ should be considered compactification-independent even though it apparently involves fields associated with the internal manifold $M$, namely the field $J^{++}_C$. The reason for this is that the internal part is always modeled by a topologically twisted ${\cal N}=4$ superconformal field theory in the hybrid formulation. Therefore the symmetry currents generating the ${\cal N}=4$ superconformal algebra, particularly $J^{++}_C$, are always present for any choice of the compactification manifold that is consistent with six-dimensional space-time supersymmetry. Of course, in principle there might be additional fields that only exists for certain choices of the internal manifold contributing to the physical spectrum; \textit{e.g.}\ if the theory is compactified on the four-torus, the topologically twisted SCFT has two complex fermions of conformal weight zero, which might also give rise to physical states. But these should not be considered compactification-independent, since they are specific to toroidal compactifications, in contrast to the superconformal currents. Following this philosophy, we generalise the vertex operator in (\ref{vo_massless}) to include the first affine excitations of the WZW currents and of the scalar bosons $\rho$, $\sigma$ and $H$. Thus our ansatz to describe compactification-independent states on the first massive level is
\begin{equation}
V^+ \equiv \no{\phi_{1,0} e^{\rho+iH}} + \no{(\phi_{2,1} + \partial \rho\, \phi^\rho_{2,1} +i\partial \sigma\, \phi^\sigma_{2,1} + i\partial H\, \phi^H_{2,1}) e^{2 \rho + i \sigma+iH}} \,.
\end{equation}
Here $\phi_{2,1}$ is a vertex operator of the $\rm PSL(2|2)$ WZW model of zero conformal weight while $\phi_{1,0}$, $\phi^\rho_{2,1}$, $\phi^\sigma_{2,1}$ and $\phi^H_{2,1}$ are associated to affine ground states of the $\rm PSL(2|2)$ WZW model of conformal weight $-1$. It is easily checked that $V^+$ has  $U(1)$-charge one and vanishing conformal weight, as required by the physical state conditions in the hybrid formulation.

\subsubsection{The $\tilde{G}^+_0$ condition}

We first check what constraints on the $\phi$'s are imposed by the $\tilde{G}^+_0 = 0$ condition. Recalling that $\tilde{G}^+ = e^{\rho+iH}$ we can determine the residue of the OPE with each summand. The evaluation of the residues, although straightforward, is in some cases cumbersome. The reader interested in the technical details of the calculation is referred to appendix \ref{hybrid_OPE}, where the rather involved calculation of the residue in (\ref{OPE_example}) is presented in detail. The relevant residues for the $\tilde{G}^+_0$ condition are
\begin{align}
\res\,  e^{\rho+iH}(z)\, \no{\phi_{1,0} e^{\rho+iH}}(w)  &= 0 \,, \\[0.25cm]
\res\, e^{\rho+iH}(z)\, \no{\phi_{2,1} e^{2\rho+i\sigma+iH}}(w) &= 0 \,, \\[0.25cm]
\res\,e^{\rho+iH}(z)\, \no{\phi^\sigma_{2,1} i\partial \sigma e^{2 \rho + i \sigma+iH}}(w) &= 0 \,, \\[0.25cm]
\res\, e^{\rho+iH}(z)\, \no{\phi^\rho_{2,1} \partial \rho e^{2 \rho + i \sigma+iH}}(w) &= -\no{\phi^\rho_{2,1} e^{3 \rho+i\sigma+2iH}}(w) \,, \\[0.25cm]
\res\, e^{\rho+iH}(z)\, \no{\phi^\rho_{2,1} \partial \rho e^{2 \rho + i \sigma+iH}}(w) &= 2 \no{\phi^H_{2,1} e^{3 \rho + i\sigma+2iH}}(w) \,.
\end{align}
Hence we conclude that physical states have to satisfy
\begin{equation}
\phi^\rho_{2,1} = 2\, \phi^H_{2,1} \label{connection_rho_H}
\end{equation}
in order for the full residue to vanish. In the following, this relation is imposed on $V^+$.

\subsubsection{The $G^+_0$ condition}

We now turn to the second kernel condition $G^+_0 = 0$. The calculation can be simplified by noting that the sum of all normal ordered products that are proportional to some exponential $e^{m \rho + i n \sigma + iH}$ in the first order pole of the OPE has to vanish independently. Hence we will consider them separately:

\para{Terms proportional to $e^{2 \rho + 2 i \sigma + i H}$: } 

 Note that  normal ordered products proportional to $e^{2 \rho + 2 i \sigma + i H}$ only appear in the OPE of $\no{e^{i \sigma} {\cal T}}$ with summands of $V^+$ proportional to $e^{2 \rho + i \sigma + i H}$. Thus the residue of this OPE has to vanish separately. One obtains
\begin{align}
\res \no{e^{i \sigma} {\cal T}}(z)  &\no{\phi_{2,1} e^{2 \rho + i \sigma+iH}}(w)  \nonumber \\[0.25cm]
&= \no{i \partial \sigma \{L_1 \phi_{2,1}\} e^{2\rho + 2i\sigma+iH}}(w) \,, \label{calT_with_21} \\[0.25cm]
\res \no{e^{i \sigma} {\cal T}}(z) & \no{ i\partial \sigma \phi^\sigma_{2,1} e^{2 \rho + i \sigma+iH}}(w) \nonumber \\[0.25cm]
&= -\no{\left(\{L_{-1} \phi^\sigma_{2,1}\}+2\partial(\rho+i\sigma) \phi^\sigma_{2,1} \right) e^{2\rho + 2i\sigma+iH}}(w) \,, \label{OPE_example} \\[0.25cm]
\res \no{e^{i \sigma} {\cal T}}(z) &\no{\partial(2\rho+iH) \phi^H_{2,1} e^{2 \rho + i \sigma+iH}}(w) \nonumber \\[0.25cm]
&= -\no{\partial(2\rho+iH) \phi^H_{2,1} e^{2 \rho +2 i \sigma+iH}}(w) \,.
\end{align}
The full residue therefore vanishes if
\begin{align}
\phi^\sigma_{2,1} = \phi^H_{2,1} &= 0 \qquad \text{and} \label{cond_on_ghosts} \\
 L_1 \phi_{2,1} &= 0 \,.
 \end{align} 
Condition (\ref{cond_on_ghosts}) together with (\ref{connection_rho_H}) tells us that all ghost excitations at the first affine level are unphysical. As a consequence, we only have to look at the space of states of the WZW-model. The second condition further restricts the physical sector to include only Virasoro primaries up to the first level. Of course, the field $\phi_{1,0}$ has not yet been restricted in any way except for being an affine ground state; it is therefore a Virasoro primary by construction. Hence we reduced physical vertex operators to be of the form
\begin{equation}
V^+ = \no{\phi_{1,0} e^{\rho+iH}} + \no{\phi_{2,1} e^{2 \rho + i \sigma+iH}}
\end{equation}
with Virasoro primaries $\phi_{1,0}$ and $\phi_{2,1}$.

\para{Terms proportional to $e^{\rho + i \sigma + i H}$:}

The next step is to look at those terms of the residue proportional to $e^{\rho+i\sigma+iH}$. These may come from the OPE of $\no{e^{i\sigma}{\cal T}}$ with terms in $\no{\phi_{1,0} e^{\rho+iH}}$ and from the OPE of $\no{e^{-\rho} Q}$ with $\no{\phi_{2,1} e^{2\rho+i\sigma+iH}}$. Their residues read
\begin{align}
\res \no{e^{i\sigma}{\cal T}}(z) &\no{\phi_{1,0} e^{\rho+iH}}(w) \nonumber \\[0.25cm]
&= \no{\left(\{L_{-1} \phi_{1,0}\}+i\partial \sigma \{L_0 \phi_{1,0}\} + \partial(\rho+i\sigma) \phi_{1,0}\right) e^{\rho+i\sigma+iH}}(w) \nonumber \\[0.25cm]
&= \no{\left(\{L_{-1} \phi_{1,0}\} + \partial \rho \phi_{1,0}\right)e^{\rho+i\sigma+iH}}(w) \,, \label{calT_with_10} \\[0.25cm]
\res \no{e^{-\rho} Q}(z) &\no{\phi_{2,1} e^{2\rho+i\sigma+iH}}(w) \nonumber \\[0.25cm]
&= \no{ \left( \{Q_0 \phi_{2,1}\} - \partial \rho \{Q_1 \phi_{2,1} \}\right) e^{\rho+i\sigma+iH}}(w) \label{Q_with_21} \,,
\end{align}
where we used that $L_0 \phi_{1,0} = -\phi_{1,0}$. Again, we demand the full residue to vanish. We find that $\phi_{1,0}$ is determined by $\phi_{2,1}$,
\begin{equation}
\phi_{1,0} = Q_1 \phi_{2,1} \,, \label{phi10}
\end{equation}
and that $\phi_{2,1}$ is subject to the constraint,
\begin{equation}
Q_0 \phi_{2,1} + L_{-1} \phi_{1,0} = (Q_0 + L_{-1} Q_1) \phi_{2,1} = 0 \,.
\end{equation}
Note that (\ref{phi10}) implies that $\phi_{1,0}$ does not carry physical degrees of freedom even though it is non-vanishing. Thus it is enough to know all the physical degrees of freedom contained in $\phi_{2,1}$. Apart from the condition that it has to be a Virasoro primary with respect to the WZW-model, it also lies in the kernel of $Q_0 + L_{-1} Q_1$. 

\para{Terms proportional to $e^{iH}$:}

Finally, we take a look at terms of the residue uncharged with respect to $\rho$ as well as $\sigma$. In other words, we consider terms proportional to $e^{iH}$ originating from the OPE of $\no{e^{-\rho} Q}$ with $\no{\phi_{1,0} e^{\rho+iH}}$ and the OPE of $\no{e^{-2\rho - i \sigma} P}$ with $\no{\phi_{2,1} e^{2 \rho + i \sigma+iH}}$,
\begin{align}
\res \no{e^{-\rho} Q}(z)&\no{\phi_{1,0} e^{\rho+iH}}(w) \nonumber \\[0.25cm]
&= \no{\left(\{Q_{-1} \phi_{1,0}\} - \partial \rho \{Q_0 \phi_{1,0}\}\right)e^{iH} }(w) \,, \label{Q_with_10} \\[0.25cm]
\res \no{e^{-2\rho - i \sigma} P}(z)& \no{\phi_{2,1} e^{2 \rho + i \sigma+iH}}(w) \nonumber \\[0.25cm] 
&= \no{\left(\{P_0 \phi_{2,1}\} - \partial(2 \rho + i \sigma) \{P_1 \phi_{2,1}\}\right) e^{iH}}(w) \label{P_with_21} \,.
\end{align}
From those and the identification in (\ref{phi10}) we see that $\phi_{2,1}$ is also subject to the constraints
\begin{equation}
(P_0 - Q_{-1} Q_1) \phi_{2,1} =  Q_0 Q_1 \phi_{2,1} = P_1 \phi_{2,1} = 0 \,.
\end{equation}

\para{Terms proportional to $e^{- \rho - i \sigma + i H}$:}

There is still one OPE left to consider, namely the OPE of $\no{e^{-2 \rho - i \sigma} P}$ with $\no{\phi_{1,0} e^{\rho + i H}}$. Its residue is
\begin{align}
\res \no{e^{-2 \rho - i \sigma} P}(z)&\no{\phi_{1,0} e^{\rho + i H}}(w) \nonumber \\[0.25cm]
&= \no{\left(\{P_{-1} \phi_{1,0}\} - \partial(2\rho + i \sigma) \{P_0 \phi_{1,0}\} \right) e^{-\rho-i \sigma+iH}}(w) \,. 
\end{align}
So we must demand
\begin{equation}
P_{-1} Q_1 \phi_{2,1} = P_0 Q_1 \phi_{2,1} = 0 \,.
\end{equation}
in order to exploit the kernel condition completely.

Let us summarise our results so far. We have seen that the compactification-independent physical spectrum of the hybrid string can be identified within the $\rm PSL(2|2)$ WZW model alone, at least up to the first level. The physical hybrid string states within the full WZW-spectrum are subject to the following kernel conditions
\begin{equation}
L_1 \phi  = (Q_0 + L_{-1} Q_1) \phi = (P_0 - Q_{-1} Q_1) \phi = Q_0 Q_1 \phi = P_{-1} Q_1 \phi = P_0 Q_1 \phi = P_1 \phi = 0 \,. \label{hybrid_kernel_conditions}
\end{equation}
The interpretation of the first constraint is simply that physical hybrid string states have to be Virasoro primary as already said above. We will see later that apart from the restriction to be Virasoro primaries, the only  significant condition to impose is the second one in (\ref{hybrid_kernel_conditions}), namely $(Q_0 + L_{-1} Q_1) \phi = 0$. All other kernel conditions are then automatically satisfied, at least at the first level.

\subsubsection{Gauge Degrees of Freedom}

Having found a set of kernel conditions on physical states, eq. (\ref{hybrid_kernel_conditions}), we may now investigate which of these are gauge trivial. In particular, we are interested in the gauge freedoms of the WZW vertex operator $\phi_{2,1}$ as it is the only one carrying physical degrees of freedom. 

Recall that gauge trivial states in the hybrid formulation are of the form $\tilde{G}^+_0 G^+_0 \Lambda^-$. In order for that vertex operator to be of the same form as $\phi_{2,1} e^{2 \rho + i \sigma + iH}$, a sensible ansatz for $\Lambda^-$ is
\begin{equation}
\Lambda^- = \Lambda_{1,0} e^{\rho} + \Lambda_{2,1} e^{2\rho + i\sigma} \,,
\end{equation}
where $\Lambda_{1,0}$ and $\Lambda_{2,1}$ are vertex operators associated to WZW states that lie in an affine Kac module of lowest conformal weight $-1$. In particular, $\Lambda_{1,0}$ has conformal weight $-1$, \textit{i.e.} it corresponds to an affine ground state, and $\Lambda_{2,1}$ has vanishing conformal weight so that $\Lambda^-$ has conformal weight zero as well. As before, the terms in $\tilde{G}^+_0 G^+_0 \Lambda^-$ can again be distinguished by the exponential in the $\rho\sigma$-ghosts, which can be considered individually. 

\para{Terms proportional to $e^{3 \rho + 2 i \sigma + i H}$: }

First, we look at terms proportional to $e^{3\rho+2i\sigma+iH}$. They only arise from the OPE of $\no{e^{i\sigma} {\cal T}}$ with $\no{\Lambda_{2,1} e^{2\rho+i\sigma}}$. From eq. (\ref{calT_with_21}) we can immediately extract that
\begin{align}
\res \no{e^{i \sigma} {\cal T}}(z)  &\no{\Lambda_{2,1} e^{2 \rho + i \sigma}}(w)  \nonumber \\[0.25cm]
&= \no{i \partial \sigma \{L_1 \Lambda_{2,1}\} e^{2\rho + 2i\sigma+iH}}(w) \,.  \label{Gauge_Vir_Primary}
\end{align}
Applying $\tilde{G}^-_0$ to this first order pole, we obtain the required terms proportional to  $e^{3\rho+2i\sigma+iH}$. Since such terms do not appear in $V^+$, we have to require that these terms vanish. So the gauge parameter $\Lambda_{2,1}$ has to be annihilated by $L_1$. In other words, it has to be a Virasoro primary.

\para{Terms proportional to $e^{2 \rho + i \sigma + i H}$: }

Next, we take a look at the terms proportional to  $e^{2\rho+i\sigma+iH}$. These may be regarded as the most important ones because they describe the gauge freedom of $\phi_{2,1}$. Similar as before, they arise from the OPE of $\no{e^{i\sigma}{\cal T}}$ with terms in $\no{\Lambda_{1,0} e^{\rho+iH}}$ and from the OPE of $\no{e^{-\rho} Q}$ with $\no{\Lambda_{2,1} e^{2\rho+i\sigma+iH}}$. From eq. (\ref{calT_with_10}) and (\ref{Q_with_21}) we know that
\begin{align}
\res \no{e^{i\sigma}{\cal T}}(z) &\no{\Lambda_{1,0} e^{\rho}}(w) + \no{e^{-\rho} Q}(z) \no{\Lambda_{2,1} e^{2\rho+i\sigma}}(w) \nonumber \\[0.25cm]
&= \no{ \left(\{L_{-1} \Lambda_{1,0}\}  + \{Q_0 \Lambda_{2,1}\} +  (\Lambda_{1,0} - \{Q_1 \Lambda_{2,1} \} )\partial \rho \right) e^{\rho+i\sigma}}(w) \,. \end{align}
We have to demand that $\Lambda_{1,0} = Q_1 \Lambda_{2,1}$ since otherwise $\tilde{G}^+_0$ applied to the first order pole above yields a vertex operator involving $\rho$-ghost excitations. Substituting $\Lambda_{1,0}$, we then obtain a gauge freedom of $\phi_{2,1}$,
\begin{equation}
\phi_{2,1} \quad \sim \quad \phi_{2,1} + (Q_0 + L_{-1} Q_1) \Lambda_{2,1} \,.
\end{equation}
Note that the operator that acts on the gauge parameter $\Lambda_{2,1}$ is exactly the deformation of $Q_0$ that appears in the kernel conditions (\ref{hybrid_kernel_conditions}) suggesting a cohomological description.

\para{Terms proportional to $e^{\rho + i H}$: }

The terms proportional to $e^{\rho+iH}$ in $\tilde{G}^+_0 G^+_0 \Lambda^-$ arise from the OPE of $\no{e^{-\rho} Q}$ with $\no{\Lambda_{1,0} e^{\rho}}$ and the OPE of $\no{e^{-2\rho - i \sigma} P}$ with $\no{\Lambda_{2,1} e^{2 \rho + i \sigma}}$. Using (\ref{Q_with_10}) and (\ref{P_with_21}), we obtain
\begin{align}
&\res \no{e^{-\rho} Q}(z)\no{\Lambda_{1,0} e^{\rho}}(w) - \no{e^{-2\rho - i \sigma} P}(z) \no{\Lambda_{2,1} e^{2 \rho + i \sigma}}(w) \nonumber \\[0.25cm] 
&\hspace{0.25cm}= \no{\{Q_{-1} \Lambda_{1,0}\}-\{P_0 \Lambda_{2,1}\} + \partial(2 \rho + i \sigma) \{P_1 \Lambda_{2,1}\} - \partial \rho \{Q_0 \Lambda_{1,0}\} }(w) \equiv {\cal F}(w)  \,. 
\end{align}
We temporarily denote this first order term by ${\cal F}(w)$. Note that it involves no exponential in the $\rho\sigma$ ghosts at all. Hence when we determine the first order pole with $\tilde{G}^+ = e^{\rho+iH}$, the first two terms vanish. In the end, it simplifies to
\begin{equation}
\res\, e^{\rho+iH}(z) {\cal F}(w) = \no{\bigl( 2 \{P_1 \Lambda_{2,1}\} - \{Q_0 \Lambda_{1,0}\}\bigr) e^{\rho+iH}} \,.
\end{equation}
Therefore, the gauge parameter $\Lambda_{2,1}$ induces a change of $\phi_{1,0}$ as
\begin{equation}
\phi_{1,0} \sim \phi_{1,0} + (2 P_1 - Q_0 Q_1) \Lambda_{2,1} \label{phi_change} \,.
\end{equation}
But $\phi_{1,0}$ carries no independent physical degrees of freedom since it descends from $\phi_{2,1}$ by applying $Q_1$. Now using that $L_1 \Lambda_{2,1} = 0$ because of (\ref{Gauge_Vir_Primary}) and $\Lambda_{2,1} \in \affK^{(1)}(\lambda)$ with $\lambda$ chosen such that $L_0 \Lambda_{2,1} = 0$, one can show that 
\begin{equation}
(2 P_1 - Q_0 Q_1) \Lambda_{2,1}  = Q_1(Q_0 + L_{-1} Q_1) \Lambda_{2,1}
\end{equation}
and hence the change in $\phi_{1,0}$ in (\ref{phi_change}) descends from the gauge freedom of $\phi_{2,1}$ as one would have hoped. We conclude that physical states $\phi$ are only well defined up to a gauge freedom,
\begin{equation}
\phi \quad \sim \quad \phi + (Q_0 + L_{-1} Q_1) \Lambda \qquad \text{with} \quad L_1 \Lambda = 0 \,. \label{hybrid_gauge_dof}
\end{equation}
Thus we have arrived at a classification of the physical spectrum at the first level in terms of the algebraic structure of $\psl$.

\subsection{Cohomological Description of Physical String States}

So far we have worked in the hybrid formulation and reduced the physical state conditions to algebraic requirements on states in the WZW factor. We will now have a closer look at the algebraic constraints (\ref{hybrid_kernel_conditions}) and (\ref{hybrid_gauge_dof}). Recall that WZW vertex operators $\phi_{2,1}(z)$ and $\Lambda_{2,1}(z)$ correspond to states of the WZW model at the first level of vanishing conformal weight, hence from a representation theoretic point of view, we are looking for $\phi, \Lambda \in \affK^{(1)}(\lambda)$ with $\lambda$ chosen such that $L_0 \phi = L_0 \Lambda = 0$. This is exactly the case if $C_2(\lambda) = -2k$. In the following we will assume that $\lambda$ has been chosen in this manner. 

Let us first consider some of the kernel conditions (\ref{hybrid_kernel_conditions}) in detail. The condition $L_1 \phi = 0$ implies that physical string states are Virasoro primaries. We can define a projection $\Pi^{(1)}$ onto the kernel of $L_1$, by
\begin{equation}
\Pi^{(1)} = \mathds{1} + \tfrac{1}{2} L_{-1}L_1 \,. \label{projection_1}
\end{equation}
Note that we are working at the first level with the ground states having conformal weight~$-1$. Furthermore, any affine ground state $\ket{\lambda} \in \affK^{(1)}(\lambda)$ satisfies $\Pi^{(1)} L_{-1} \ket{\lambda} = 0$. This implies that the first level decomposes as $\widehat{\cal K}^{(1)}(\lambda) = \ker L_1 \oplus \mathrm{im}\, L_{-1}$. Thus restricting $\affK^{(1)}(\lambda)$ to the $\g$-submodule of Virasoro primaries is, morally speaking, equivalent to identifying the submodule of Virasoro descendants and removing it from the decomposition in (\ref{first_affine_lvl}). Note that $L_{-1}$ applied to the ground state representation $\widehat{\cal K}^{(0)}(\lambda)\bigr|_{\mathfrak{g}} \simeq {\cal K}(\lambda)$ yields a copy of it at the first level,
\begin{equation}
\mathrm{im}\, L_{-1} \, \simeq \, {\cal K}(\lambda) \quad \subset \quad \widehat{\cal K}^{(1)}(\lambda) \,.
\end{equation}
Therefore the subspace of Virasoro primaries at the first level decomposes into $\mathfrak{g}$-represen\-tations as (recall that the subscript denotes the grading of the cyclic state of the respective Kac-module)
\begin{align}
\ker_{\widehat{\cal K}^{(1)}(\lambda)} L_1 \Bigr|_\mathfrak{g} \simeq & \phantom{\oplus} {\cal K}_0(\lambda) \oplus {\cal K}_0(\lambda^{++}) \oplus {\cal K}_0(\lambda^{--}) \oplus {\cal K}_0(\lambda_{++}) \oplus {\cal K}_0(\lambda_{--}) \nonumber \\
& \oplus \bigoplus_{g = \pm 1} \Bigl({\cal K}_g(\lambda^+_+) \oplus {\cal K}_g(\lambda^+_-) \oplus {\cal K}_g(\lambda^-_+) \oplus {\cal K}_g(\lambda^-_-)\Bigr) \,. \label{level_one_Vir_primaries}
\end{align}

The next condition of interest is the deformation of $Q_0$ in (\ref{hybrid_kernel_conditions}), namely $Q_0 + L_{-1} Q_1$. In order to get some insight into the meaning of this operator we can proceed as follows. Having the projection (\ref{projection_1}) onto $\ker L_1$ at hand, we can project $Q_0$ onto $\ker L_1$ simply by multiplying $\Pi^{(1)}$ on both sides. The projected operator is denoted as $Q^\Pi_0 \equiv \Pi^{(1)} Q_0 \Pi^{(1)}$. Since the Virasoro modes $L_n$ commute with the horizontal subalgebra of $\affpsl$, the projection $\Pi^{(1)}$ commutes with it as well. Thus we can use (\ref{comm_rel_QS}) to calculate the commutation relation
\begin{align}
\left[Q^\Pi_0, S^{\alpha\beta}_{+,0}\right] &= \left[\Pi^{(1)} Q_0 \Pi^{(1)}, S^{\alpha\beta}_{+,0}\right] = \Pi^{(1)}\, \left[Q_0, S^{\alpha\beta}_{+,0}\right]\,  \Pi^{(1)} \nonumber \\ 
& \hspace{2cm} =  -i\sqrt{k}\,  (S^{\alpha\beta}_{-,0} L^\Pi_0+ \Pi^{(1)} L_{-1} S^{\alpha\beta}_{-,1} \Pi^{(1)} + \Pi^{(1)}S^{\alpha\beta}_{-,-1} L_1\Pi^{(1)}) \,, \label{commutator_Q_S2}
\end{align}
where $L^\Pi_n \equiv  \Pi^{(1)} L_n  \Pi^{(1)}$ is a shorthand notation for the projected Virasoro modes. Because $ \Pi^{(1)} L_{-1} = 0$ when applied to affine ground states in $\affK^{(0)}(\lambda)$ and $L_1 \Pi^{(1)} = 0$ on $\affK^{(1)}(\lambda)$, the last two terms in (\ref{commutator_Q_S2}) vanish. Hence
\begin{align}
\left[Q^\Pi_0, S^{\alpha\beta}_{+,0}\right] =  -i\sqrt{k}\, S^{\alpha\beta}_{-,0} L^\Pi_0 \,.
\end{align}
It is easy to check that the zero mode $L_0$ commutes with $\Pi^{(1)}$ using the commutation relations of Virasoro modes. Since this implies that $L^\Pi_0 = \Pi^{(1)} L_0$, the above commutator vanishes when applied to the kernel of $L_0$, \textit{i.e.}\ to states of zero conformal weight. Because $\lambda$ is chosen such that $\affK^{(1)}(\lambda)$ is annihilated by $L_0$, \textit{i.e.}\ $C_2(\lambda) = -2k$, we obtain
\begin{align}
\left[Q^\Pi_0, S^{\alpha\beta}_{+,0}\right]=  0 \qquad \text{on }\ \affK^{(1)}(\lambda) \,.
\end{align}
Since both $Q_0$ and $\Pi^{(1)}$ commute with the $\mathfrak{g}^{(0)} \oplus \mathfrak{g}_{-1}$ as well, we conclude that $Q_0^\Pi$ commutes with the full horizontal algebra $\mathfrak{g}$ of $\widehat{\mathfrak{g}}$ and hence induces a $\mathfrak{g}$-homomorphism on $\affK^{(1)}(\lambda)$. 

As soon as we have imposed the physical state condition $L_1 \phi = 0$, $\Pi^{(1)}$ obviously acts on the remaining states like the identity. The action of $Q_0^\Pi$ can be evaluated on $\phi \in \ker L_1$,
\begin{equation}
Q^\Pi_0 \phi = \Pi^{(1)} Q_0 \Pi^{(1)}\, \phi = (\mathds{1} + \tfrac{1}{2} L_{-1}L_1) Q_0 \phi = (Q_0 + L_{-1} Q_1) \phi \,, \label{Q_Pi_simplification}
\end{equation}
where we used that $[L_1, Q_0] = 2\, Q_1$ holds on $\affK^{(1)}(\lambda)$ (cf.\ eq.\ (\ref{L_Q_comm})). But the last expression in (\ref{Q_Pi_simplification}) coincides exactly with the operator in (\ref{hybrid_kernel_conditions}). Thus we have obtained a nice algebraic interpretation of one of the operators appearing in (\ref{hybrid_kernel_conditions}). Namely, it is simply $Q_0$ appropriately corrected such that it induces a $\g$-homomorphism on $\ker L_1 \subset \affK^{(1)}(\lambda)$.

Before we continue analysing the kernel conditions, a discussion of the gauge degrees of freedom (\ref{hybrid_gauge_dof}) is in order. The gauge parameter $\Lambda$ has to be Virasoro primary as well and hence according to our discussion before, the gauge freedom can be equivalently written as
\begin{equation}
\phi \quad \sim \quad \phi + Q^\Pi_0 \Lambda \qquad \text{with} \quad L_1 \Lambda = 0 \,. 
\end{equation}
In other words, physical states are only defined up to states in the image of the $\g$-homomorphism induced by $Q^\Pi_0$. Clearly, this is only well-defined if $(Q^\Pi_0)^2 = 0$. Indeed, note that $(Q^\Pi_0)^2$ is a $\g$-homomorphism of grading $-4$. But the decomposition in (\ref{first_affine_lvl}) implies that no nontrivial $\g$-homomorphism mapping $\affK^{(1)}(\lambda) \rightarrow \affK^{(1)}(\lambda)$ of grading $-4$ exists. Hence the $\g$-homomorphism induced by $Q^\Pi_0$ is nilpotent and the $Q^\Pi_0$-cohomology is well-defined on the submodule of Virasoro primaries.

The spectrum of Virasoro primaries has been given in (\ref{level_one_Vir_primaries}). Since $Q^\Pi_0$ is an operator of grading $-2$, we can immediately state that all Kac-modules of zero grading will contribute to the $Q^\Pi_0$-cohomology. Unfortunately, just by considering the grading of $Q^\Pi_0$, we cannot make any statement on the remaining Kac-modules in (\ref{level_one_Vir_primaries}) because the $\g$-homomorphism
\begin{equation}
Q^\Pi_0: \qquad {\cal K}_{+1}(\lambda^\alpha_\beta) \longrightarrow {\cal K}_{-1}(\lambda^\alpha_\beta) \,, \qquad \alpha,\beta = \pm\,,
\end{equation}
might be nontrivial. Using the explicit realisation of $Q_0$ in (\ref{explicit_Q_1}) and (\ref{explicit_Q_2}), one can show that the induced homomorphism from $\K_{+1}(\lambda^\alpha_\beta)$ to $\K_{-1}(\lambda^\alpha_\beta)$ is indeed nontrivial for any combination of $\alpha$ and $\beta$. In this sense, the action of $Q^\Pi_0$ is maximal. Hence we conclude that
\begin{equation}
H_{Q^\Pi_0}\left(\ker L_1\right)\Bigr|_\mathfrak{g} \simeq {\cal K}_0(\lambda) \oplus {\cal K}_0(\lambda^{++}) \oplus {\cal K}_0(\lambda^{--}) \oplus {\cal K}_0(\lambda_{++}) \oplus {\cal K}_0(\lambda_{--}) \,. \label{Q_cohom_1}
\end{equation}

We now turn our attention to the other conditions in (\ref{hybrid_kernel_conditions}). We want to show that they are all automatically satisfied once $\phi$ is taken to be an element of the $Q^\Pi_0$-cohomology. We begin by showing $(P_0 - Q_{-1} Q_1) \phi = 0$. As for $Q_0$, the zero-mode $P_0$ commutes with $\mathfrak{g}_{-1} \oplus \mathfrak{g}^{(0)}$, but a priori does not commute with $\mathfrak{g}_{+1}$. However, taking into account the correction term $-Q_{-1}Q_1$, we find that at the first level the commutation relation
\begin{align}
[P_0 - &Q_{-1} Q_{1},S^{\alpha\beta}_{+,0}] \nonumber \\
&= i \sqrt{k} \left( S^{\alpha\beta}_{-,0}(Q_0 + Q_{-1} L_{1} + L_{-1} Q_1) + (Q_{-1} S^{\alpha\beta}_{-,1} + S^{\alpha\beta}_{-,-1} Q_1) L_0\right) 
\end{align}
holds. When acting on states in $\ker L_1 \subset \affK^{(1)}(\lambda)$, which are annihilated by both $L_0$ and $L_1$, the above commutation relations simplify to
\begin{equation}
[P_0 - Q_{-1} Q_{1}, S^{\alpha\beta}_{+,0}] = i \sqrt{k}\,  S^{\alpha\beta}_{-,0}(Q_0 + L_{-1} Q_1) =  i \sqrt{k}\,  S^{\alpha\beta}_{-,0}\, Q^\Pi_0  \,,
\end{equation}
where we have made use of (\ref{Q_Pi_simplification}). So after restricting to the subspace of Virasoro primaries within $\widehat{\cal K}^{(1)}(\lambda)$ and imposing the physical state condition $Q^\Pi_0 \phi = 0$, the deformed operator $P_0 - Q_{-1} Q_{1}$ induces a $\mathfrak{g}$-homomorphism  of grading $-4$ on the resulting subspace. But since the $Q^\Pi_0$-kernel only involves Kac-modules of grading $0$ and $-1$, it is clear that $P_0 - Q_{-1} Q_{1}$ annihilates all states in that kernel. Thus the $(P_0 - Q_{-1}Q_1)$-kernel condition is trivially satisfied once we reduced the physical subspace to be a part of the $Q^\Pi_0$-kernel. Hence the $(P_0 - Q_{-1} Q_{1})$-kernel condition may be discarded.

The two operators $P_{-1} Q_1$ and $P_0 Q_1$ commute with the bosonic subalgebra $\mathfrak{g}^{(0)}$ of $\mathfrak{g}$ and hence induce homomorphisms of $\mathfrak{g}^{(0)}$-representations. However, these operators have grading $-6$ and hence they induce trivial homomorphisms because the $Q^\Pi_0$-kernel as well as the affine ground states $\affK^{(0)}(\lambda)$ only involve $\mathfrak{g}^{(0)}$-representations that have gradings between $0$ and $-5$.

The remaining operators in (\ref{hybrid_kernel_conditions}), $Q_0 Q_1$ and $P_1$, cannot be deduced to be trivial simply by an analysis of their gradings. However, using their explicit realisation in terms of modes, one finds that they indeed vanish on the direct summands of $\affK^{(1)}(\lambda)$ given in (\ref{Q_cohom_1}) that are neither in the kernel nor in the image of $Q^\Pi_0$.

Thus we have shown that physical string states at the first level in the $\rm PSL(2|2)$ WZW model can be described by the $Q^\Pi_0$-cohomology evaluated on the subspace of Virasoro primaries of conformal weight zero,
\begin{equation}
{\cal H}^{(1),\mathrm{PSL}}_\mathrm{phys} \simeq H_{Q^\Pi_0}\left(\ker_{\affK^{(1)}(\lambda)} L_1\right) \label{phys_PSL_1}
\end{equation}
with the weight $\lambda$ chosen such that $L_0 = 0$ is satisfied. Note that this is the same description of physical states as in the case of the massless sector. There the physical sector was given by the $Q_0$-cohomology \cite{Berkovits:1999im,Dolan:1999dc,Gaberdiel:2011vf,Troost:2011fd}. But $Q^\Pi_0$ reduces to $Q_0$ on affine ground states as they all are Virasoro primaries. Hence (\ref{phys_PSL_1}) can be considered as the natural generalisation of the description of the massless sector.

\section{Comparison with the RNS String Spectrum} \label{corr_rns}

In the previous section we have succeed to give a description of the  physical states of the hybrid formulation within the $\rm PSL(2|2)$ WZW model. Next we want to show that our result is in agreement with the spectrum one obtains for the RNS string theory on $\rm{AdS}_3 \times \rm{S}^3$ \cite{Giveon:1998ns, Evans:1998qu}. In fact, we will be considering the full string spectrum not restricted to the first massive level. This will allow us to deduce that the massive compactification-independent physical spectrum fits into representations of $\g$ at all mass levels. 

Let us assume that the NS vacuum has $\mathfrak{sl}(2) \oplus \mathfrak{sl}(2)$ quantum numbers $\lambda = (j_1, j_2)$. The generating function of physical $\mathfrak{sl}(2) \oplus \mathfrak{sl}(2)$ highest weight states of the RNS-string on $\rm{AdS}_3 \times S^3$ in the NS- and R-sector are, respectively,
\begin{align}
F^\mathrm{NS}&(x,y|q) \nonumber \\&= x^{j_1} y^{j_2} q^{\frac{1}{2k} C_2(\lambda)-\frac{1}{2}}\prod_{n \geq 1} \frac{(1+x q^{n-\frac{1}{2}})(1+x^{-1} q^{n-\frac{1}{2}})(1+y q^{n-\frac{1}{2}})(1+y^{-1} q^{n-\frac{1}{2}})}
{(1-x q^{n})(1-x^{-1} q^{n})(1-y q^{n})(1-y^{-1} q^{n})} \,, \\
F^\mathrm{R}&(x,y|q) \nonumber \\
&=  x^{j_1 + \frac{1}{2}} y^{j_2 + \frac{1}{2}} q^{\frac{1}{2k} C_2(\lambda)-\frac{1}{4}} \prod_{n \geq 1} \frac{(1+x q^{n})(1+x^{-1} q^{n-1})(1+y q^{n})(1+y^{-1} q^{n-1})}
{(1-x q^{n})(1-x^{-1} q^{n})(1-y q^{n})(1-y^{-1} q^{n})} \,.
\end{align}
These expressions can be found by thinking of $J^0$ and $K^0$ as the light cone directions. The physical state conditions in the supersymmetric $\rm SL(2) \times SU(2)$ WZW model are expected to eliminate states from the spectrum that are in one-to-one correspondence to excitations of these light cone currents; this leads to the generating functions $F^\mathrm{NS}(x,y|q)$ and $F^\mathrm{R}(x,y|q)$ above. The first few terms in the $q$-expansions have explicitly been checked to give the correct quantum numbers of the physical spectrum \cite{Gaberdiel:notes}.

In order to apply the GSO-projection later, we also have to determine the generating functions with the insertion of $(-1)^F$; they read
\begin{align}
F^{\widetilde{\mathrm{NS}}}&(x,y|q) \nonumber \\ &= x^{j_1} y^{j_2} q^{\frac{1}{2k} C_2(\lambda)-\frac{1}{2}}\prod_{n \geq 1} \frac{(1-x q^{n-\frac{1}{2}})(1-x^{-1} q^{n-\frac{1}{2}})(1-y q^{n-\frac{1}{2}})(1-y^{-1} q^{n-\frac{1}{2}})}
{(1-x q^{n})(1-x^{-1} q^{n})(1-y q^{n})(1-y^{-1} q^{n})} \,, \\
F^{\widetilde{\mathrm{R}}}&(x,y|q) \nonumber \\
 &= x^{j_1 + \frac{1}{2}} y^{j_2 + \frac{1}{2}} q^{\frac{1}{2k} C_2(\lambda)-\frac{1}{4}} \prod_{n \geq 1} \frac{(1-x q^{n})(1-x^{-1} q^{n-1})(1-y q^{n})(1-y^{-1} q^{n-1})}
{(1-x q^{n})(1-x^{-1} q^{n})(1-y q^{n})(1-y^{-1} q^{n})} \nonumber \\
&=   x^{j_1+\frac{1}{2}} y^{j_2+\frac{1}{2}} (1-x^{-1})(1-y^{-1})q^{\frac{1}{2k} C_2(\lambda)-\frac{1}{4}} \,.
\end{align}
We are interested in the spectrum of compactification-independent states, \textit{i.e.}\ the subsector of physical states that are always present independent of the choice of $M$. However, the choice of $M$ is restricted to manifolds that yield target space supersymmetry in six dimensions. The existence of supersymmetry in a six-dimensional spacetime requires the ${\cal N}=1$ superconformal symmetry of the non-linear $\sigma$-model with target space $M$ to be extended to an ${\cal N}=4$ superconformal symmetry \cite{Seiberg:1988pf,Banks:1988yz}. Hence the fields generating the ${\cal N}=4$ superconformal algebra are always present and should be considered as compactification-independent, even though they correspond to excitations on the compactification manifold. The character for some representation ${\cal D}$ of the ${\cal N}=4$ superconformal algebra is defined as
\begin{equation}
\chi^{\cal D}_{{\cal N}=4}(z|q) = \mathrm{Tr}_{\cal D} \left(q^{L_0} z^{{\cal J}_0} \right) \,,
\end{equation}
where ${\cal J}_0$ is the $U(1)$-charge of the superconformal algebra.  In \cite{Eguchi:1987wf}, the characters have been determined for large classes of representations. For the compactification-independent spectrum only the vacuum representations in the NS- and R-sector are of interest. Their respective characters are
\begin{align}
\chi^\mathrm{R}_{{\cal N}=4}(z|q) &= q^{\frac{1}{4}} \frac{i \vartheta_{10}^2(z|q)}{\vartheta_{11}(z^2|q) \eta^3(q)} \sum_{m \in \mathbb{Z}} \left(\frac{z^{4m+1}}{(1+z^{-1} q^{-m})^2} - \frac{z^{-4m-1}}{(1+z q^{-m})^2}\right) q^{2 m^2 + m} \,, \\
\chi^\mathrm{NS}_{{\cal N}=4}(z|q) &= q^{\frac{1}{4}} \frac{i \vartheta_{00}^2(z|q)}{\vartheta_{11}(z^2|q) \eta^3(q)} \sum_{m \in \mathbb{Z}} \left(\frac{z^{4m+1}}{(1+z q^{m+\frac{1}{2}})^2} - \frac{z^{-4m-1}}{(1+z^{-1} q^{m+\frac{1}{2}})^2}\right) q^{2 m^2 + m} \,.
\end{align}
The first factor containing theta functions\footnote{Our conventions for the theta functions are
\begin{align*}
\vartheta_{00}(z|q) &= \prod_{m \geq 1} (1-q^m)(1+z q^{m-\frac{1}{2}})(1+z^{-1} q^{m-\frac{1}{2}}) \,, \\ 
\vartheta_{10}(z|q) &= q^{\frac{1}{8}}(z^{\frac{1}{2}} + z^{-\frac{1}{2}}) \prod_{m \geq 1} (1-q^m)(1+z q^m)(1+z^{-1} q^m) \,, \\
\vartheta_{11}(z|q) &= i q^{\frac{1}{8}}(z^{\frac{1}{2}} - z^{-\frac{1}{2}}) \prod_{m \geq 1} (1-q^m)(1-z q^m)(1-z^{-1} q^m) \,.
\end{align*}
They satisfy the relations $\vartheta_{10}(z|q)=q^{\frac{1}{8}} z^{\frac{1}{2}} \vartheta_{00}(zq^{\frac{1}{2}}|q)$ and $\vartheta_{11}(z^2|q) = -z^2 q^{\frac{1}{2}} \vartheta_{11}(z^2 q|q)$.} is the character of the corresponding Verma module while the infinite sum encodes the singular (and redundant) vectors within that Verma module. Note that the characters of the two sectors are connected by spectral flow,
\begin{equation}
\chi^\mathrm{R}_{{\cal N}=4}(z|q) = z q^{\frac{1}{4}} \chi^\mathrm{NS}_{{\cal N}=4}(z q^{\frac{1}{2}}|q) \,.
\end{equation}
Since the supercurrents have odd $U(1)$-charge and the bosonic currents of the ${\cal N}=4$ superconformal algebra have even $U(1)$-charge, the insertion of $(-1)^F$ is easily implemented by substituting $z$ by $-z$. 
The partition sum of compactification-independent physical states in each sector is then given by the GSO-projected product of the $\rm{AdS}_3 \times S^3$ string partition sum with the respective ${\cal N}=4$ superconformal vacuum character. Thus the full compactification-independent generating function including both the NS- and the R-sector is given by
\begin{align}
Z^\mathrm{indep}(x,y|q) = & \phantom{+} \frac{1}{2} \Bigl( F^\mathrm{NS}(x,y|q)\, \chi^\mathrm{NS}_{{\cal N}=4}(z|q) - F^{\widetilde{\mathrm{NS}}}(x,y|q) \chi^\mathrm{NS}_{{\cal N}=4}(-z|q)\Bigr) \Bigr|_{z=0} \nonumber \\
& + \frac{1}{2} \Bigl( F^\mathrm{R}(x,y|q)\, \chi^\mathrm{R}_{{\cal N}=4}(z|q) - F^{\widetilde{\mathrm{R}}}(x,y|q) \chi^\mathrm{R}_{{\cal N}=4}(-z|q)\Bigr) \Bigr|_{z=0}  \,.
\end{align}
Expanding it in powers of $q$, we obtain
\begin{equation}
Z^\mathrm{indep}(x,y|q) = \sum_{n \in \mathbb{N}} {\cal A}_n(x,y) q^{n+\frac{1}{2k} C_2(\lambda)} \,.
\end{equation}
The coefficient functions  ${\cal A}_n(x,y)$ can be understood as the generating functions of the physical states at the $n$-th mass level that are highest weight states with respect to $\bg$; for the massless level, \textit{i.e.} $n = 0$, the coefficient function is for example
\begin{equation}
{\cal A}_0(x,y) = x^{j_1-1} y^{j_2 - 1}\left(x^{\frac{1}{2}}+y^{\frac{1}{2}}\right)^2 (x y+1) \,. \label{coe_fct_C0}
\end{equation}
It turns out that the coefficient function ${\cal A}_n(x,y)$ for $n \geq 1$ seems to factorise into a factor $K_{\lambda}(x,y)$ and a residual polynomial factor. This factorisation property of the coefficient functions has been checked explicitly up to mass level $n=6$. Hence we can write
\begin{equation}
{\cal A}_n(x,y) = K_{\lambda}(x,y) \sum_{r,s = 0}^{4n} (A_n)_{rs}\, x^{\frac{r}{2}-n} y^{\frac{s}{2}-n} \,,
\end{equation}
where the $A_n$ are $(4n+1) \times (4n+1)$ matrices. Since the function $x^{l_1} y^{l_2} K_{\lambda}(x,y)$ is just the branching function for ${\cal K}(\lambda + (l_1,l_2))$, we see that the massive states of the RNS-string on $\rm{AdS}_3 \times \rm{S}^3$ can be arranged into a direct sum of Kac modules with respect to $\g$. Then the matrices $A_n$ have the interpretation of encoding of the multiplicities of Kac modules at the various mass levels. The explicit form of the matrices $A_n$ with $n = 1,\ldots,6$ can be found in appendix $\ref{appB}$. For the case $n=1$ that is of primary interest to us, the relevant coefficient function is
\begin{align}
{\cal A}_1(x,y) &= K_{\lambda} (x,y) (x^{-1} + x + y^{-1} + y + 1) \,.  \nonumber \\
&= K_{\lambda} (x,y) + K_{\lambda^{++}} (x,y)+ K_{\lambda^{--}} (x,y) + K_{\lambda_{++}} (x,y) + K_{\lambda_{--}} (x,y) \label{C_1}
\end{align}
The associated matrix $A_1$ can be easily extracted from this expression and it is found to be 
\begin{equation}
A_1 =  \left( \begin{array}{ccccc}
 0 & 0 & 1 & 0 & 0 \\
 0 & 0 & 0 & 0 & 0 \\
 1 & 0 & 1 & 0 & 1 \\
 0 & 0 & 0 & 0 & 0 \\
 0 & 0 & 1 & 0 & 0 
 \end{array}
\right) \,.
\end{equation}
Hence the physical compactification-independent RNS string states on $\rm{AdS}_3 \times \rm{S}^3$ can be uniquely arranged in the following direct sum of $\psl$ representations:
\begin{equation}
{\cal H}^{(1),\mathrm{RNS}}_\mathrm{phys} \simeq {\cal K}(\lambda) \oplus {\cal K}(\lambda^{++}) \oplus {\cal K}(\lambda^{--}) \oplus {\cal K}(\lambda_{++}) \oplus {\cal K}(\lambda_{--}) \,. \label{phys_RNS_1}
\end{equation}

Eq. (\ref{phys_RNS_1}) should now be compared with our result obtained in the hybrid formulation in (\ref{phys_PSL_1}). Indeed, the spectra on first level agree nicely,
\begin{equation}
{\cal H}^{(1),\mathrm{PSL}}_\mathrm{phys} \quad = \quad {\cal H}^{(1),\mathrm{RNS}}_\mathrm{phys} \,.
\end{equation}
This is a good consistency check of our analysis.

\section{A Conjecture and its Confirmation at the Second Affine Level} \label{conjecture}

With our previous results in mind, it seems natural to expect that the cohomological characterisation of physical string states in the $\rm PSL(2|2)$ WZW model generalises to all mass levels. In particular, at level $n$, we conjecture that the space of physical states is given by
\begin{equation}
{\cal H}^{(n)}_\mathrm{phys} \quad \simeq \quad H_{Q^\Pi_0}\left( \mathrm{Vir}\, \affK^{(n)}(\lambda) \right) \,,
\end{equation}
where $\lambda$ is chosen such that $C_2(\lambda) = -n$, and $\mathrm{Vir}\, \affK^{(n)}(\lambda)$ denotes the $\g$-submodule of Virasoro primaries within $\affK^{(n)}(\lambda)$, \textit{i.e.}
\begin{equation}
\mathrm{Vir}\, \affK^{(n)}(\lambda)  \quad \equiv \quad \left\{ \phi \in \affK^{(n)}(\lambda) \, \Big|\ L_m \phi = 0  \quad \forall m \geq 1\right\} \,.
\end{equation}
Furthermore $\Pi$ is the projection onto the subspace of Virasoro primaries. We assume that any state can be uniquely decomposed into a Virasoro primary and a Virasoro descendant. In other words, our assumption is that $\affK(\lambda)$ can be written as a direct sum of the space of Virasoro primaries and the space of Virasoro descendants. This seems reasonable since any Verma module with respect to the $c=-2$ Virasoro algebra of highest weight $h \leq -1$ is irreducible \cite{Feigin:1988se}. Hence the space of Virasoro descendants does not contain Virasoro primaries since they would generate a subrepresentation. In the rest of this section, we confirm the conjectured characterisation of physical compactification-independent string states in the $\rm PSL(2|2)$ WZW model at the second level.

We start by defining the modified branching function of $\affK(\lambda)$, which is
\begin{align}
\chi_{\affK_0(\lambda)}(x,y,u|q) &= \mathrm{Tr}^{(0)}_{\affK_0(\lambda)} \left(u^\varrho x^{J^0} y^{K^0} q^{L_0-\frac{1}{2k}C_2(\lambda)}\right)  \,,
\end{align}
where the trace $\mathrm{Tr}^{(0)}$ is only taken over highest weight states with respect to the bosonic subalgebra $\g^{(0)}$ of the horizontal subalgebra $\g$. We have furthermore introduced a chemical potential $u$ that keeps track of the grading of these states. $C_2(\lambda)$ is the value of the quadratic Casimir evaluated on the ground state representation $\K(\lambda)$; by subtracting $\frac{1}{2k} C_2(\lambda)$ from the $L_0$-eigenvalue the branching function has no poles in $q$. Evaluating the trace yields
\begin{align}
\chi_{\affK_0(\lambda)}(x,y,u|q)  &= x^{j_1} y^{j_2} \prod_{n \geq 1} \frac{\prod_{\alpha,\beta = \pm 1} (1+u x^{\frac{\alpha}{2}}y^{\frac{\beta}{2}} q^n) (1+u^{-1} x^{\frac{\alpha}{2}} y^{\frac{\beta}{2}} q^{n-1})}
 {(1-x^{-1} q^n)(1-x q^n)(1-y^{-1} q^n)(1-y q^n)(1-q^n)^2} \,.
\end{align}
Loosely speaking, the factors in the numerator correspond to the fermionic generators of $\g$ while the factors in the denominator are associated to the bosonic ones. The first step on the way to determine the claimed space of physical states is then to identify the subspace of Virasoro primaries.  By our assumption that $\affK(\lambda)$ decomposes into Virasoro primaries and descendants, it is sufficient to eliminate the Virasoro Verma module generated by the affine ground states. This is most easily achieved by multiplying the character $\chi_{\affK(\lambda)}$ with $q^{-\frac{1}{24}} \eta(q)$, the inverse of the Virasoro vacuum character. The spectrum of Kac modules that are Virasoro primary can be extracted form the resulting character by expanding it as
\begin{equation}
 \chi_{\mathrm{Vir}\,\affK(\lambda)}(x,y,u|q) = q^{-\frac{1}{24}} \eta(q) \, \chi_{\affK(\lambda)}(x,y,u|q) = K_{\lambda}(x,y) \sum_{g \in \mathbb{Z} \atop n \in \mathbb{N}} \sum_{r,s = 0}^{4n} \left(D^n_g\right)_{rs}\, x^{\frac{r}{2}-n} y^{\frac{s}{2}-n} u^g q^n \,.
\end{equation}
Similar to the matrices $C_n$ defined below eq. (\ref{coe_fct_C0}), the matrices $D^n_g$ encode the multiplicities of the various Kac-modules  of grading $g$ at level $n$ that are Virasoro primaries. Clearly, $D^0_0 = 1$ and $D^0_g = 0$ for all $g \neq 0$. For $n=1$ we get
\begin{equation}
D^1_0 = \left(\begin{array}{ccccc} 
0 & 0 & 1 & 0 & 0 \\
0 & 0 & 0 & 0 & 0 \\
1 & 0 & 1 & 0 & 1 \\
0 & 0 & 0 & 0 & 0 \\
0 & 0 & 1 & 0 & 0
\end{array} \right) \,, \qquad 
D^1_{\pm 1} = \left(\begin{array}{ccccc} 
0 & 0 & 0 & 0 & 0 \\
0 & 1 & 0 & 1 & 0 \\
0 & 0 & 0 & 0 & 0 \\
0 & 1 & 0 & 1 & 0 \\
0 & 0 & 0 & 0 & 0
\end{array} \right) \,,
\end{equation}
which agrees with (\ref{level_one_Vir_primaries}). We have seen that all Kac modules of grading $+1$, which are counted by $D^1_{+1}$, are mapped by $Q^\Pi_0$ to Kac modules of grading $-1$, which are counted by $D^1_{-1}$. Since $D^1_{-1} = D^1_{+1}$, none of them survive in the cohomology. Thus $D^1_0$ encodes the physical string spectrum at first level in agreement with (\ref{Q_cohom_1}) and (\ref{phys_PSL_1}).

Expanding $\chi_{\mathrm{Vir}\,\affK(\lambda)}$ to second order in $q$, we obtain the following set of matrices:
\begin{equation*}
 D^2_0 = \left(
\begin{array}{ccccccccc}
 0 & 0 & 0 & 0 & 1 & 0 & 0 & 0 & 0 \\
 0 & 0 & 0 & 0 & 0 & 0 & 0 & 0 & 0 \\
 0 & 0 & 2 & 0 & 4 & 0 & 2 & 0 & 0 \\
 0 & 0 & 0 & 0 & 0 & 0 & 0 & 0 & 0 \\
 1 & 0 & 4 & 0 & 8 & 0 & 4 & 0 & 1 \\
 0 & 0 & 0 & 0 & 0 & 0 & 0 & 0 & 0 \\
 0 & 0 & 2 & 0 & 4 & 0 & 2 & 0 & 0 \\
 0 & 0 & 0 & 0 & 0 & 0 & 0 & 0 & 0 \\
 0 & 0 & 0 & 0 & 1 & 0 & 0 & 0 & 0
\end{array}
\right) \,,
\,
 D^2_{\pm 1} = \left(
\begin{array}{ccccccccc}
 0 & 0 & 0 & 0 & 0 & 0 & 0 & 0 & 0 \\
 0 & 0 & 0 & 1 & 0 & 1 & 0 & 0 & 0 \\
 0 & 0 & 0 & 0 & 0 & 0 & 0 & 0 & 0 \\
 0 & 1 & 0 & 4 & 0 & 4 & 0 & 1 & 0 \\
 0 & 0 & 0 & 0 & 0 & 0 & 0 & 0 & 0 \\
 0 & 1 & 0 & 4 & 0 & 4 & 0 & 1 & 0 \\
 0 & 0 & 0 & 0 & 0 & 0 & 0 & 0 & 0 \\
 0 & 0 & 0 & 1 & 0 & 1 & 0 & 0 & 0 \\
 0 & 0 & 0 & 0 & 0 & 0 & 0 & 0 & 0
\end{array}
\right) \,, \,
 D^2_{\pm 2} = \left(
\begin{array}{ccccccccc}
 0 & 0 & 0 & 0 & 0 & 0 & 0 & 0 & 0 \\
 0 & 0 & 0 & 0 & 0 & 0 & 0 & 0 & 0 \\
 0 & 0 & 0 & 0 & 1 & 0 & 0 & 0 & 0 \\
 0 & 0 & 0 & 0 & 0 & 0 & 0 & 0 & 0 \\
 0 & 0 & 1 & 0 & 2 & 0 & 1 & 0 & 0 \\
 0 & 0 & 0 & 0 & 0 & 0 & 0 & 0 & 0 \\
 0 & 0 & 0 & 0 & 1 & 0 & 0 & 0 & 0 \\
 0 & 0 & 0 & 0 & 0 & 0 & 0 & 0 & 0 \\
 0 & 0 & 0 & 0 & 0 & 0 & 0 & 0 & 0
\end{array}
\right) \,.
\end{equation*}
We will need the projection operator onto the Virasoro primaries up to second level. It is realised in terms of Virasoro modes as
\begin{equation}
\Pi^{(2)} = \mathds{1} + \tfrac{1}{2} L_{-1} L_1 + \tfrac{1}{10} L_{-1}^2 L_1^2 + \tfrac{1}{15} L_{-2} L_2 + \tfrac{1}{30} \left(L_{-1}^2 L_2 + L_{-2} L_1^2 \right) \,. \label{projection_2}
\end{equation}
Note that 
\begin{equation}
\left(\Pi^{(2)}\right)^2 = \Pi^{(2)} \quad + \quad (\text{terms annihilating states in } \affK^{(n)}(\lambda) \text{ for } n = 0,1,2) \,,
\end{equation}
and that $\left(\mathds{1}-\Pi^{(2)}\right)$ clearly maps to Virasoro descendants. Furthermore, eq. (\ref{explicit_Q_1}) and (\ref{explicit_Q_2}) give an explicit realisation of $Q_0$ as an infinite sum of products of $\affg$-modes. When applied to states at the second level, this infinite sum truncates to a finite number of terms. Thus, using (\ref{projection_2}), we can work with an expression for $Q^\Pi_0$ that has only finitely many terms. After identifying the cyclic states of the Kac modules at second level $\affK^{(2)}(\lambda)$ of $\affK(\lambda)$, it is possible to check with this realisation of $Q^\Pi_0$ that
\begin{itemize}

\item the homomorphism induced by $Q^\Pi_0$ satisfies $\left(Q^\Pi_0\right)^2=0$, and hence the \\ \mbox{$Q^\Pi_0$-cohomology} is well defined,

\item no Kac module of grading $+2$ lies in the kernel of $Q^\Pi_0$, \textit{i.e.}\ the Kac modules in $D^2_{+2}$ do not contribute to the cohomology and neither does their image in $D^2_0$,

\item every Kac module of grading $-2$ lies in the image of $Q^\Pi_0$, \textit{i.e.}\ the Kac modules in $D^2_{-2}$ do not contribute to the cohomology and neither does their preimage in $D^2_0$\,.

\end{itemize}
The Kac modules of odd grading are less intuitive as the action of $Q^\Pi_0$ is not maximal in the sense that not every Kac module in $D^2_{+1}$ is mapped to one in $D^2_{-1}$. In fact, by a brute force calculation using the explicit realisations of the Kac modules as well as the operators $Q_0$ and $\Pi$, one finds that four Kac-modules in $D^2_{+1}$ are in the kernel of $Q^\Pi_0$. As a consequence, the same set of Kac modules does not lie in the image of $Q^\Pi_0$ in $D^2_{-1}$. In our matrix notation the states that survive in cohomology can be encoded as
\begin{equation*}
\widetilde{D}^2_{\pm 1} \equiv \left(
\begin{array}{ccccccccc}
 0 & 0 & 0 & 0 & 0 & 0 & 0 & 0 & 0 \\
 0 & 0 & 0 & 0 & 0 & 0 & 0 & 0 & 0 \\
 0 & 0 & 0 & 0 & 0 & 0 & 0 & 0 & 0 \\
 0 & 0 & 0 & 1 & 0 & 1 & 0 & 0 & 0 \\
 0 & 0 & 0 & 0 & 0 & 0 & 0 & 0 & 0 \\
 0 & 0 & 0 & 1 & 0 & 1 & 0 & 0 & 0 \\
 0 & 0 & 0 & 0 & 0 & 0 & 0 & 0 & 0 \\
 0 & 0 & 0 & 0 & 0 & 0 & 0 & 0 & 0 \\
 0 & 0 & 0 & 0 & 0 & 0 & 0 & 0 & 0
\end{array}
\right) \,, 
\end{equation*}
and hence the spectrum of Kac module surviving in cohomology is encoded in the matrix
\begin{equation}
D^2_0 - D^2_{-2} - D^2_{+2} + \widetilde{D}^2_{-1}+ \widetilde{D}^2_{+1} = \left(
\begin{array}{ccccccccc}
 0 & 0 & 0 & 0 & 1 & 0 & 0 & 0 & 0 \\
 0 & 0 & 0 & 0 & 0 & 0 & 0 & 0 & 0 \\
 0 & 0 & 2 & 0 & 2 & 0 & 2 & 0 & 0 \\
 0 & 0 & 0 & 2 & 0 & 2 & 0 & 0 & 0 \\
 1 & 0 & 2 & 0 & 4 & 0 & 2 & 0 & 1 \\
 0 & 0 & 0 & 2 & 0 & 2 & 0 & 0 & 0 \\
 0 & 0 & 2 & 0 & 2 & 0 & 2 & 0 & 0 \\
 0 & 0 & 0 & 0 & 0 & 0 & 0 & 0 & 0 \\
 0 & 0 & 0 & 0 & 1 & 0 & 0 & 0 & 0
\end{array}
\right) \,.
\end{equation}
This matrix is the same as the matrix $A_2$ in appendix \ref{appB}, which tells us the multiplicities and $\g$-quantum numbers of physical Kac modules in the RNS formulation. We therefore conclude that our conjecture holds at the second level.

\section{Conclusions} \label{conclusion}

In this paper, we have succeeded in evaluating the physical state constraints of the hybrid formulation at the first level. We have found a surprisingly simple and elegant description of the physical compactification-independent string spectrum in the context of the $\rm PSL(2|2)$ WZW model. In particular, we have shown that the physical string states can be identified within the $\rm PSL(2|2)$ WZW model by taking an appropriate cohomology of the full WZW spectrum at the first massive level. This cohomological description of the physical states reduces to the one known for the supergravity spectrum when applied to the affine ground states. Hence it seems to be the appropriate generalisation of the description of compactification-independent physical string states in the massless sector. This strongly suggests that the description of physical states we found can be further extended to describe also the physical states at any mass level. A possible generalisation to all mass levels was conjectured and found to agree at the second level with the on shell RNS string spectrum.

Within the program of understanding the string theory side of the AdS$_3$/CFT$_2$ correspondence, the natural next step 
would be to investigate how our description of physical string states changes if the
$\rm PSL(2|2)$ WZW model is marginally deformed such that RR flux is included to the string background. Even though
no RNS description of string theory exists in that case, from the point of view of the hybrid formulation
this is equivalent to leaving the WZW point in moduli space. In general, calculations away from the WZW point
are difficult to perform due to the fact that the currents are not holomorphic anymore, which forbids the
application of complex analytic tools (see \cite{Ashok:2009jw, Benichou:2010rk} for recent advances in this direction). But since there exists a non-renormalisation theorem specific to the $\rm PSL(2|2)$
WZW model \cite{Bershadsky:1999hk}, one may expect that it is possible to keep track of how the description of
physical states changes along any path in the moduli space that starts at the WZW point. 
\section*{Acknowledgements}

The author is grateful to M.~R.~Gaberdiel for support and instructive discussions. Furthermore, he would like thank C.~Candu, M.~Kelm, I.~Kirsch and C.~Vollenweider for helpful discussions. This work was partially supported by the Swiss National Science Foundation.

\newpage

\appendix

\section{Bases and Commutator Relations of $\affpsl$} \label{app:A}

The Lie superalgebra $\g = \mathfrak{psl}(2|2)$ can be decomposed as
\begin{equation}
 \g = \g_{+1} \oplus \g^{(0)} \oplus \g_{-1} \,,
\end{equation}
where $\g^{(0)}$ is the bosonic subalgebra and $\g^{(1)} = \g_{+1} \oplus \g_{-1}$ consists of
the fermionic generators. The bosonic generators are denoted by $K^{ab}$ with 
$\mathfrak{so}(4)$-indices $a,b$ and the fermionic generators are denoted by
$S^a_\alpha \in \g_{\alpha}$ where $\alpha = \pm$. Hence the index $\alpha$ corresponds
to the $\Z$-grading $\varrho$ as explained in sect. \ref{intro_psl}. For later use, we
also define  $\varepsilon_{\alpha\beta}$ as 
\begin{equation}
 \varepsilon_{+-} = -\varepsilon_{-+} = 1 \,, \qquad \varepsilon_{++} = \varepsilon_{--} = 0 \,.
\end{equation}
In the basis used in \cite{Dolan:1999dc, Troost:2011fd}, the commutation relations of the affine version $\affpsl$ read
\begin{align*}
[K^{ab}_m,K^{cd}_n] &=  i \left( \delta^{ac}K^{bd}_{m+n}-\delta^{bc}K^{ad}_{m+n}-\delta^{ad}K^{bc}_{m+n}+\delta^{bd}K^{ac}_{m+n}\right) + n k \delta_{n+m} \varepsilon^{abcd} \\[2mm]
[K^{ab}_m,S^c_{\gamma,n}] &=  i \left( \delta^{ac} S^b_{\gamma,m+n} - \delta^{bc} S^a_{\gamma,m+n} \right) \\[2mm]
[ S^a_{\alpha,m},S^b_{\beta,n}] & =  \tfrac{i}{2}\,\varepsilon_{\alpha\beta}\, \varepsilon^{abcd} K^{cd}_{m+n} + nk\varepsilon_{\alpha\beta} \delta^{ab} \delta_{m+n}  \ ,
\end{align*}
We will mainly just write upper indices for readability and impose the summation convention if the same index appears twice as upper (lower) index as well. An appropriate basis change can be made by defining \cite{Gotz:2006qp}
\begin{align*}
  J^0_n &= \tfrac{1}{2}\left(K^{12}_n+K^{34}_n\right) & K^0_n &= \tfrac{1}{2}\left(K^{12}_n-K^{34}_n\right)  \\[2mm]
  J^\pm_n &= \tfrac{1}{2} \left(K^{14}_n+K^{23}_n\pm iK^{24}_n\mp iK^{13}_n\right)  &K^\pm_n &= \tfrac{1}{2}\left(-K^{14}_n+K^{23}_n\mp iK^{24}_n\mp iK^{13}_n\right)  \\[2mm]
  S_{\alpha,n}^{\pm\pm} &= S_{\alpha,n}^1\pm iS_{\alpha,n}^2 & S_{\alpha,n}^{\pm\mp} &= S_{\alpha,n}^3\pm iS_{\alpha,n}^4 \ ,
\end{align*}
for which the commutation relations are explicitly given by
\begin{align*}
   [J_m^0,J_n^0]& = -\frac{k}{2}\,m\,\delta_{m+n} \,, &
   [K_m^0,K_n^0]& = \frac{k}{2}\,m\,\delta_{m+n} \,, \\[2mm]
   [J_m^0,J_n^\pm]& = \pm\,J_{n+m}^\pm \,, &
   [K_m^0,K_n^\pm]& = \pm\,K_{n+m}^\pm \,, \\[2mm]
  [J_m^+,J_n^-]& = 2J_{n+m}^0 -mk\,\delta_{m+n} \,, &
   [K_m^+,K_n^-]& = 2K_{n+m}^0 +mk\,\delta_{m+n} \,, \\[2mm]
   [J_m^0,S^{\pm\pm}_{\alpha,n}]& = \pm\frac{1}{2}\,S^{\pm\pm}_{\alpha,m+n} \,, &
   [J_m^0,S^{\pm\mp}_{\alpha,n}]& =
\pm\frac{1}{2}\,S^{\pm\mp}_{\alpha,m+n} \,, \\[2mm]
   [K_m^0,S^{\pm\pm}_{\alpha,n}]& = \pm\frac{1}{2}\,S^{\pm\pm}_{\alpha,m+n} \,, &
   [K_m^0,S^{\pm\mp}_{\alpha,n}]& =
\mp\frac{1}{2}\,S^{\pm\mp}_{\alpha,m+n} \,, \\[2mm]
   [J_m^\pm,S_{\alpha,n}^{\mp\mp}]& = \pm S^{\pm\mp}_{\alpha,m+n} \,, &
   [J_m^\pm,S_{\alpha,n}^{\mp\pm}]& = \mp S^{\pm\pm}_{\alpha,m+n} \,, \\[2mm]
   [K_m^\pm,S_{\alpha,n}^{\mp\mp}]& = \pm S_{\alpha,m+n}^{\mp\pm} \,, &
   [K_m^\pm,S^{\pm\mp}_{\alpha,n}]& = \mp S^{\pm\pm}_{\alpha,m+n} \,, \\[2mm] 
   \{S^{\pm\pm}_{\alpha,m},S^{\pm\mp}_{\beta,n}\}& = \mp 2\epsilon_{\alpha\beta}\,J_{n+m}^\pm \,,&
   \{S^{\pm\pm}_{\alpha,m},S_{2\beta,n}^\mp\}& = \pm 2\epsilon_{\alpha\beta}\,K_{n+m}^\pm \,,
\end{align*}
\begin{align*}
   \{S_{\alpha,m}^{++},S_{\beta,n}^{--}\}& = %
2\epsilon_{\alpha\beta}\bigl(J_{n+m}
^0-K_{n+m}^0\bigr)- 2\,mk\,\epsilon_{\alpha\beta}\,\delta_{m+n} \,, \\[2mm]
   \{S_{\alpha,m}^{+-},S_{\beta,n}^{-+}\}& = %
2\epsilon_{\alpha\beta}\bigl(J_{n+m} ^0+K_{n+m}^0\bigr) -
2\,mk\,\epsilon_{\alpha\beta}\,\delta_{m+n} \, .
\end{align*}
Of course, the commutation relations of $\psl$ can be extracted from the affine commutation relations by looking at the zero modes only, \textit{i.e.}\ the horizontal subalgebra.
The quadratic Casimir operator of $\psl$ is
\begin{equation}
 C_2 = C^\text{bos}_2 + C^\text{fer}_2
\end{equation}
with
\begin{align}
C_2^\text{bos} &= - 2 (J^0)^2 - (J^+ J^- + J^- J^+) + 2 (K^0)^2 + (K^+ K^- + K^- K^+) \\
C_2^\text{fer} &= \frac{\epsilon^{\alpha\beta}}{2}\sum_{m=1}^{2}\left(S^+_{m\alpha}S^-_{m\beta}+S^-_{m\alpha}S^+_{m\beta}\right) 
= \sum_{m=1}^{2}\left(S^+_{m-}S^-_{m+}+S^-_{m-}S^+_{m+}\right) \   . \label{Casfer}
\end{align}
The operator $C_2^\text{fer}$ is the only bilinear in the fermionic generators that commutes with 
the bosonic subalgebra $\bg$. Evaluated on an irreducible representation of highest weight $\lambda = (j_1, j_2)$, \textit{e.g.}\ the Kac module $\K(\lambda)$, it takes the value
\begin{equation}
C_2(\lambda) \equiv -2\,j_1(j_1+1) + 2\, j_2(j_2+1) \,.
\end{equation}

\section{The Hybrid Vertex Operator $Q$ and its Properties} \label{app_Q}

Recall that whenever $\mathfrak{so}(4)$-indices $a,b,c,\ldots$ appear twice within a term, summation is understood. The field $Q$ as defined in (\ref{def_Q}),
\begin{equation}
Q \equiv \frac{1}{2\sqrt{k}} \left[ \no{K_{ab} \no{S^a_- S^b_-}} + 4 i \no{S^a_- \partial S^a_-} \right] \,,
\end{equation}
can be easily written in terms of modes using general results on the mode expansion of the normal ordered product of two vertex operators \cite{Borcherds:1983sq,Gaberdiel:2000qn}. Specifically, if $\psi(z)$ and $\chi(w)$ are two vertex operators, the $n$-th mode of the product equals
\begin{equation}
(\psi \chi)_n = \sum_{L \geq 0} \left( \psi_{-h_\psi - L} \chi_{n+h_\psi + L} + \varepsilon\, \chi_{n+h_\psi-1-L} \psi_{-h_\psi+1+L} \right) \,, \label{no_expression}
\end{equation}
where $h_\psi$ is the conformal weight of $\psi$. Here $\varepsilon = -1$ if both $\psi$ and $\chi$ are fermionic and $\varepsilon = +1$ otherwise. With this result it is straightforward to write down a mode expression for the $n$-th mode of $Q$. For the individual summands of $Q$ we get
\begin{align}
(K^{ab} (S^a_- S^b_-))_n &= \sum_{m,l \geq 0} \Bigl[K^{ab}_{-m-1} S^a_{-,-l-1} S^b_{-,l+m+n+2} + K^{ab}_{-m-1} S^a_{-,m+n+1-l} S^b_{-,l} \nonumber \\
& \hspace{3cm} + S^a_{-,-l-1} S^b_{-,n-m+l+1} K^{ab}_m + S^a_{-,n-m-l} S^b_{-,l} K^{ab}_m\Bigr] \nonumber \\
&= \sum_{m,l \in \mathbb{Z}} K^{ab}_{-m} S^a_{-,-l} S^b_{-,n+m+l} \,, \label{explicit_Q_1}
\end{align}
where $\sum_{l \in \mathbb{Z}} [K^{ab}_m, S^a_{-l} S^b_{n+l}] = 0$ was used, and
\begin{align}
(S^a \partial S^a)_n &= \sum_{m \geq 0} \left( (-n-m-2) S^a_{-,-m-1} S^a_{-,n+m+1} + (n-m+1) S^a_{-,n-m} S^a_{-,m}  \right) \nonumber \\
& \hspace{1.5cm} = \sum_{m \in \mathbb{Z}} (m-n-1) \, S^a_{-,m} S^a_{-,n-m} = \sum_{m \in \mathbb{Z}} (-m-1) \, S^a_{-,n-m} S^a_{-,m}  \,. \label{explicit_Q_2} 
\end{align}
An important property of $Q$ is its commutation relation with operators in $\g_{+1}$,
\begin{equation}
[Q_0, S^a_{+,0}] = -i \sqrt{k} \no{S^a_- T^\mathrm{WZW}}_0 \,, \label{comm_rel_Q_g+1}
\end{equation}
where $\alpha,\beta = \pm 1$. This relation can be proven by noting that
\begin{equation}
\left[S^a_{+,0}, Q(w)\right] = \left\{S^a_{+,0}\, Q_{-3} \Omega \right\}(w) \,,
\end{equation}
where $\Omega$ is the conformal vacuum annihilated by all non-negative modes of the WZW-currents. The state associated to the field $Q$ by the operator-state-correspondence is
\begin{align}
Q_{-3} \Omega &=  \frac{1}{2\sqrt{k}} \left[\no{K^{ab} \no{S^a_- S^b_-}}_{-3} \Omega + 4i \no{S^a \partial S^a}_{-3} \Omega\right] \nonumber \\
&=  \frac{1}{2\sqrt{k}}\left(K^{ab}_{-1} S^a_{-1} S^b_{-1} + 4 i S^a_{-1} S^a_{-2} \right)\Omega \,.
\end{align}
It is straightforward to evaluate the commutators of $S^c_{+,0}$ with $Q_{-3}$. First, we consider the commutator
\begin{equation}
\left[S^c_{+,0}, K^{ab}_{-1} S^a_{-,-1} S^b_{-,-1}\right] = 2i\, S^c_{-,-1} S^a_{+,-1} S^a_{-,-1} - \tfrac{i}{2} \varepsilon^{cbef} S^a_{-,-1} \{K^{ab}_{-1}, K^{ef}_{-1}\} + \Xi^c_{-3} \,. \label{app_comm_1}
\end{equation}
The auxiliary operator $\Xi^c_{-3}$ collects all residual terms that involve elementary commutators of the modes of the WZW currents and its explicit form will be discussed later. The anticommutator $\{K^{ab}_{-1}, K^{ef}_{-1}\}$ can be considered as a tensor $T^{abef}$ with the symmetry properties
\begin{equation}
T^{abef} = -T^{baef} = -T^{abfe} = T^{efab} \,.
\end{equation}
For such tensor, the following identity holds
\begin{equation}
\varepsilon^{cabe} T^{adbe} = -\tfrac{1}{4} \varepsilon^{abef}T^{abef} \delta^d_c \,.
\end{equation}
So the commutator (\ref{app_comm_1}) simplifies to
\begin{equation}
\left[S^c_{+,0}, K^{ab}_{-1} S^a_{-,-1} S^b_{-,-1}\right] = -i\, S^c_{-,-1} \left( \tfrac{1}{4} \varepsilon^{abef} K^{ab}_{-1}, K^{ef}_{-1} - 2\, S^a_{+,-1} S^a_{-,-1}\right) + \Xi^c_{-3} \,.
\end{equation}
The term in brackets should be recognised as the state that corresponds to the Sugawara energy momentum tensor $L^\mathrm{WZW}_{-2} \Omega$, multiplied by $2 k$. So in order for (\ref{comm_rel_Q_g+1}) to be valid, one would hope that the the residual term $\Xi^c_{-3}$ is canceled by the commutator of $S^c_{+,0}$ with the remaining term in $Q_{-3} \Omega$. Evaluating $\Xi^c_{-3}$, we obtain
\begin{align}
\Xi^c_{-3} &= -\tfrac{i}{2} \varepsilon^{cbef} S^a_{-1} \left[K^{ab}_{-1}, K^{ef}_{-1}\right] - 2i\,\left\{S^a_{+,-1}, S^c_{-,-1}\right\} S^a_{-,-1}
 \nonumber \\
& \hspace{1cm} -i\,  \varepsilon^{cbef}  \left[K^{ab}_{-1}, S^a_{-,-1}\right] K^{ef}_{-1} + \tfrac{i}{2}  \varepsilon^{caef} K^{ab}_{1} \left[K^{ef}_{-1},S^b_{-,-1}\right] \nonumber \\
&= - 2\, \varepsilon^{caef} S^a_{-,-1} K^{ef}_{-2} + 2\, \varepsilon^{caef} K^{ef}_{-1} S^a_{-,-2} \,,
\end{align}
which is indeed canceled by the commutator
\begin{align}
\left[S^c_{+,0},  4 i S^a_{-1} S^a_{-2} \right] = -2\, \varepsilon^{caef} K^{ef}_{-1} S^a_{-,-2} + 2\, \varepsilon^{caef} S^a_{-,-1} K^{ef}_{-2} = - \Xi^c_{-3} \,.
\end{align}
Therefore acting with $S^e_{+,0}$ on $Q_{-3} \Omega$ yields
\begin{align}
S^e_{+,0} Q_{-3} \Omega &=  \frac{i}{2\sqrt{k}} S^e_{-,-1} \left(-\tfrac{1}{4} \varepsilon_{abcd} K^{ab}_{-1} K^{cd}_{-1} + \, S^a_{+,-1} S^a_{-,-1} - S^a_{-,-1} S^a_{+,-1}\right)\Omega \nonumber \\
&= i \sqrt{k} \, S^e_{-,-1} L_{-2} \Omega = i \sqrt{k} \no{S^e_- T^\mathrm{WZW}}_{-3} \Omega \,,
\end{align}
and we conclude that
\begin{equation}
\left[S^a_{+,0}, Q(w)\right] = i \sqrt{k} \no{S^a_- T^\mathrm{WZW}}(w) \,,
\end{equation}
which in particular implies the commutation relation (\ref{comm_rel_Q_g+1}).

\noindent The right-hand side of (\ref{comm_rel_Q_g+1}) can be expanded in modes by using (\ref{no_expression}),
\begin{equation}
 \no{S^a_- T^\mathrm{WZW}}_0 =\sum_{n \in \mathbb{Z}} :L_{-n} S^a_{-,n}: \,. \label{comm_rel_Q_g+1_modes}
 \end{equation}
Recall that $:\cdot:$ refers to creation-annihilation-ordering. Note that there is no ordering ambiguity for the zero-modes since $L_0$ commutes with $S^{\alpha\beta}_{-,0}$. From (\ref{comm_rel_Q_g+1_modes}) we see that the commutation relation (\ref{comm_rel_Q_g+1}) implies the particularly interesting fact that when acting on the subspace of Virasoro primary states of conformal weight zero, $Q_0$ commutes with $\mathfrak{g}_{+1}$ up to terms that are Virasoro descendants. 

From the explicit realisations of the normal ordered products (\ref{explicit_Q_1}) and (\ref{explicit_Q_2}) in terms of modes, we can draw further conclusions. From (\ref{explicit_Q_1}) and the commutation relations $[L_{n'}, {\cal J}_n] = -n {\cal J}_{n+n'}$, where ${\cal J}$ may denote any current of the $\rm PSL(2|2)$ WZW model, one obtains
\begin{equation}
\left[ L_{n'}, \no{K_{ab}\no{S^a_- S^b_-}}_n\right] = (2 n' - n) \no{K_{ab}\no{S^a_- S^b_-}}_{n + n'} \,. \label{temp}
\end{equation}
Therefore $\no{K_{ab}\no{S^a_- S^b_-}}$ is a Virasoro primary field of conformal weight 3. The additional term (\ref{explicit_Q_2}) of $Q$ spoils this property. Indeed, we find
\begin{equation}
\left[ L_{n'}, \no{S^a_- \partial S^b_-}_n\right] = \sum_{m \in \Z} \left(\frac{n'(1-n'+2m)}{m+1} - n\right) (-m-1) S^a_{-,n+n'-m} S^a_{-,m} \,.
\end{equation}
However, if restricted to $\affK^{(1)}(\lambda)$, \textit{i.e.}\ states at the first affine level, and choosing $n=0$ and $n'=1$, the above commutator simplifies to
\begin{equation}
\left[ L_1, \no{S^a_- \partial S^b_-}_0\right]\Bigr|_{\affK^{(1)}(\lambda)} = \sum_{m =0}^1 (-2m)\, S^a_{-,1-m} S^a_{-,m} = -2 S^a_{-,0} S^a_{-,1} = 2\, \no{S^a_- \partial S^b_-}_1 \,.
\end{equation}
Combining this with (\ref{temp}), we obtain
\begin{equation}
\left[ L_1, Q_0 \right] \Bigr|_{\affK^{(1)}(\lambda)} = 2 Q_1 \label{L_Q_comm}
\end{equation}
as claimed below (\ref{Q_Pi_simplification}).

\section{Residues in the Hybrid Formulation} \label{hybrid_OPE}

Although the determination of the first order poles in section \ref{ev_hybrid_constraints} is mostly straightforward, in some cases the analysis is somewhat more involved. In this appendix, we explain the calculation of the residue for one representative case, namely (\ref{OPE_example}), in order to present the technical tools used to extract the residue and to give an impression of the basic philosophy of our calculation.

Recall that the OPE in (\ref{OPE_example}) is
\begin{equation}
\no{e^{i \sigma} {\cal T}}(z) \no{ i\partial \sigma \phi^\sigma_{2,1} e^{2 \rho + i \sigma+iH}}(w) \,.
\end{equation}
Since the $\rho\sigma$-deformed energy momentum tensor ${\cal T}(z)$ is a sum of vertex operators, we can consider each summand individually. First, let us check the OPE
\begin{equation}
\no{e^{i \sigma} T^\mathrm{WZW}}(z) \no{ i\partial \sigma \phi^\sigma_{2,1} e^{2 \rho + i \sigma+iH}}(w) \,.
\end{equation}
Using the elementary OPEs
\begin{equation}
T^\mathrm{WZW}(z) \phi^\sigma_{2,1}(w) = \sum_{l \in \mathbb{Z}} \{L_{-2-l} \phi^\sigma_{2,1}\} (z-w)^l
\end{equation}
as well as
\begin{align}
\no{e^{i \sigma}}(z) \no{ i\partial \sigma e^{2 \rho + i \sigma+iH}}(w) &= \sum_{d \geq 0} (z-w)^{d+1} \frac{1}{d!} \no{i \partial \sigma (\partial^d e^{i \sigma}) e^{2 \rho + i \sigma + i H}} \nonumber \\
& \hspace{1cm} - \sum_{d \geq 0} (z-w)^{d} \frac{1}{d!} \no{(\partial^d e^{i \sigma}) e^{2 \rho + i \sigma + i H}} \,,
\end{align}
we deduce the following OPE
\begin{align}
\no{e^{i \sigma} T^\mathrm{WZW}}(z) &\no{ i\partial \sigma \phi^\sigma_{2,1} e^{2 \rho + i \sigma+iH}}(w) \nonumber \\
&= \sum_{l \in \mathbb{Z} \atop d \geq 0} \biggl( (z-w)^{l+d+1} \frac{1}{d!} \no{\{L_{-2-l} \phi^\sigma_{2,1}\} i \partial \sigma (\partial^d e^{i \sigma}) e^{2 \rho + i \sigma + i H}} \nonumber \\
&\hspace{1cm} - (z-w)^{l+d} \frac{1}{d!} \no{\{L_{-2-l} \phi^\sigma_{2,1}\} (\partial^d e^{i \sigma}) e^{2 \rho + i \sigma + i H}} \biggr)\,.
\end{align}
In this argument it is important to realise that the $\rho\sigma$-ghosts are free and thus have non-singular OPEs with vertex operators of the $\rm PSL(2|2)$ WZW model. The first order pole can be determined by extracting the summand with $l+d+1=-1$ and $l+d=-1$, respectively. This yields
\begin{align}
\res\,\no{e^{i \sigma} T^\mathrm{WZW}}(z) &\no{ i\partial \sigma \phi^\sigma_{2,1} e^{2 \rho + i \sigma+iH}}(w) \nonumber \\
&= \sum_{d \geq 0} \biggl(  \frac{1}{d!} \no{\{L_{d} \phi^\sigma_{2,1}\} i \partial \sigma (\partial^d e^{i \sigma}) e^{2 \rho + i \sigma + i H}} \nonumber \\
&\hspace{1cm} - \frac{1}{d!} \no{\{L_{d-1} \phi^\sigma_{2,1}\} (\partial^d e^{i \sigma}) e^{2 \rho + i \sigma + i H}} \biggr)\,.
\end{align}
We can now use the properties of the vertex operator $\phi^\sigma_{2,1}$ to truncate the summation. Note that by construction of the ansatz $V^+$, the vertex operator $\phi^\sigma_{2,1}$ is an affine ground state and hence is annihilated by all positive modes. Thus the residue simplifies to
\begin{align}
\res\,\no{e^{i \sigma} T^\mathrm{WZW}}(z) &\no{ i\partial \sigma \phi^\sigma_{2,1} e^{2 \rho + i \sigma+iH}}(w) \nonumber \\
&=  - \no{\{L_{-1} \phi^\sigma_{2,1}\} e^{2 \rho +2 i \sigma + i H}} \,. \label{calc_OPE_1}
\end{align}
Next, we look at the deformation term. Useful elementary OPEs are
\begin{align}
\no{e^{i\sigma} (\partial \rho)^2}(z) \no{i \partial \sigma e^{2\rho+i\sigma+iH}}(w) &\sim \tfrac{-4}{(z-w)^2} \no{e^{2\rho+2i\sigma+iH}}(w) \nonumber \\
&\hspace{2cm} + \tfrac{4}{z-w} \no{\partial\rho e^{2\rho+2i\sigma+iH}}(w) \,, \\[0.25cm]
\no{e^{i\sigma} \partial \rho\, i \partial \sigma}(z) \no{i \partial \sigma e^{2\rho+i\sigma+iH}}(w) &\sim 0 \,, \\[0.25cm]
\no{e^{i\sigma} (i \partial \sigma)^2}(z) \no{i \partial \sigma e^{2\rho+i\sigma+iH}}(w) &\sim \tfrac{1}{(z-w)^2} \no{e^{2\rho+2i\sigma+iH}}(w) \nonumber \\
&\hspace{2cm} + \tfrac{2}{z-w} \no{i\partial\sigma e^{2\rho+2i\sigma+iH}}(w) \,, \\[0.25cm]
\no{e^{i\sigma} \partial^2 \rho}(z) \no{i \partial \sigma e^{2\rho+i\sigma+iH}}(w) &\sim \tfrac{-2}{(z-w)^2} \no{e^{2\rho+2i\sigma+iH}}(w) \\[0.25cm]
\no{e^{i\sigma}\, i \partial^2 \sigma}(z) \no{i \partial \sigma e^{2\rho+i\sigma+iH}}(w) &\sim \tfrac{-1}{(z-w)^2} \no{e^{2\rho+2i\sigma+iH}}(w) \nonumber \\
&\hspace{2cm} - \tfrac{2}{z-w} \no{i \partial\sigma e^{2\rho+2i\sigma+iH}}(w) \,.
\end{align}
Taking the appropriate sum, we obtain
\begin{align}
\no{e^{i\sigma} \Bigl(-\tfrac{1}{2} (\partial \rho + i \partial \sigma)^2 + \tfrac{1}{2} \partial^2(\rho + i \sigma) \Bigr)}(z) & \no{i \partial \sigma e^{2\rho+i\sigma+iH}}(w) \nonumber \\ 
&\sim -\tfrac{2}{z-w} \no{\partial(\rho + i\sigma) e^{2\rho+2i\sigma+iH}} \,. \label{calc_OPE_2}
\end{align}
The residue is now apparent. Combining eq.\ (\ref{calc_OPE_1}) and eq.\ (\ref{calc_OPE_2}), we obtain the full residue of the OPE of interest,
\begin{align}
\res \no{e^{i \sigma} {\cal T}}(z) &\no{ i\partial \sigma \phi^\sigma_{2,1} e^{2 \rho + i \sigma+iH}}(w) \nonumber \\
&=  - \no{\{L_{-1} \phi^\sigma_{2,1}\} e^{2 \rho +2 i \sigma + i H}}  - 2\,  \no{ \phi^\sigma_{2,1} \partial(\rho + i\sigma) e^{2\rho+2i\sigma+iH}} \,.
\end{align}
This is the residue given in (\ref{OPE_example}). The remaining residues can be determined in a similar way.

\section{Some Coefficient Functions ${\cal A}_n(x,y)$} \label{appB}

In this appendix we write the $q$-expansion of $Z^\mathrm{indep}(x,y|q)$ as
\begin{equation}
 q^{-\frac{1}{2k} C_2(\lambda)} Z^\mathrm{indep}(x,y|q) = \sum_{n \in \mathbb{N}} {\cal A}_n(x,y) q^n = {\cal A}_0(x,y) + \sum_{n \in \mathbb{N}^{\neq 0}} K_\lambda(x,y) R(x,y) q^n \,.
\end{equation}
The functions $R(x,y)$ are polynomial functions of the form
\begin{equation}
R(x,y) = \sum_{r,s = 0}^{4n} (A_n)_{rs}\, x^{\frac{r}{2}-n} y^{\frac{s}{2}-n} \,,
\end{equation}
where the $A_n$ are $(4n+1) \times (4n+1)$ matrices. These matrices tell us the multiplicities of the various Kac-modules at the different mass levels. The first matrices are given by

\begin{equation*}
 A_1 = \left(
\begin{array}{ccccc}
 0 & 0 & 1 & 0 & 0 \\
 0 & 0 & 0 & 0 & 0 \\
 1 & 0 & 1 & 0 & 1 \\
 0 & 0 & 0 & 0 & 0 \\
 0 & 0 & 1 & 0 & 0 
 \end{array}
\right) \,,
\quad
 A_2 = \left(
\begin{array}{ccccccccc}
 0 & 0 & 0 & 0 & 1 & 0 & 0 & 0 & 0 \\
 0 & 0 & 0 & 0 & 0 & 0 & 0 & 0 & 0 \\
 0 & 0 & 2 & 0 & 2 & 0 & 2 & 0 & 0 \\
 0 & 0 & 0 & 2 & 0 & 2 & 0 & 0 & 0 \\
 1 & 0 & 2 & 0 & 4 & 0 & 2 & 0 & 1 \\
 0 & 0 & 0 & 2 & 0 & 2 & 0 & 0 & 0 \\
 0 & 0 & 2 & 0 & 2 & 0 & 2 & 0 & 0 \\
 0 & 0 & 0 & 0 & 0 & 0 & 0 & 0 & 0 \\
 0 & 0 & 0 & 0 & 1 & 0 & 0 & 0 & 0
\end{array}
\right) \,,
\end{equation*}

\begin{equation*}
 A_3 = \left(
\begin{array}{ccccccccccccc}
 0 & 0 & 0 & 0 & 0 & 0 & 1 & 0 & 0 & 0 & 0 & 0 & 0 \\
 0 & 0 & 0 & 0 & 0 & 0 & 0 & 0 & 0 & 0 & 0 & 0 & 0 \\
 0 & 0 & 0 & 0 & 2 & 0 & 2 & 0 & 2 & 0 & 0 & 0 & 0 \\
 0 & 0 & 0 & 0 & 0 & 4 & 0 & 4 & 0 & 0 & 0 & 0 & 0 \\
 0 & 0 & 2 & 0 & 4 & 0 & 8 & 0 & 4 & 0 & 2 & 0 & 0 \\
 0 & 0 & 0 & 4 & 0 & 10 & 0 & 10 & 0 & 4 & 0 & 0 & 0 \\
 1 & 0 & 2 & 0 & 8 & 0 & 13 & 0 & 8 & 0 & 2 & 0 & 1 \\
 0 & 0 & 0 & 4 & 0 & 10 & 0 & 10 & 0 & 4 & 0 & 0 & 0 \\
 0 & 0 & 2 & 0 & 4 & 0 & 8 & 0 & 4 & 0 & 2 & 0 & 0 \\
 0 & 0 & 0 & 0 & 0 & 4 & 0 & 4 & 0 & 0 & 0 & 0 & 0 \\
 0 & 0 & 0 & 0 & 2 & 0 & 2 & 0 & 2 & 0 & 0 & 0 & 0 \\
 0 & 0 & 0 & 0 & 0 & 0 & 0 & 0 & 0 & 0 & 0 & 0 & 0 \\
 0 & 0 & 0 & 0 & 0 & 0 & 1 & 0 & 0 & 0 & 0 & 0 & 0
\end{array}
\right) \,,
\quad
A_4 = \left(
\begin{array}{ccccccccccccccccc}
 0 & 0 & 0 & 0 & 0 & 0 & 0 & 0 & 1 & 0 & 0 & 0 & 0 & 0 & 0 & 0 & 0 \\
 0 & 0 & 0 & 0 & 0 & 0 & 0 & 0 & 0 & 0 & 0 & 0 & 0 & 0 & 0 & 0 & 0 \\
 0 & 0 & 0 & 0 & 0 & 0 & 2 & 0 & 2 & 0 & 2 & 0 & 0 & 0 & 0 & 0 & 0 \\
 0 & 0 & 0 & 0 & 0 & 0 & 0 & 4 & 0 & 4 & 0 & 0 & 0 & 0 & 0 & 0 & 0 \\
 0 & 0 & 0 & 0 & 2 & 0 & 5 & 0 & 10 & 0 & 5 & 0 & 2 & 0 & 0 & 0 & 0 \\
 0 & 0 & 0 & 0 & 0 & 6 & 0 & 16 & 0 & 16 & 0 & 6 & 0 & 0 & 0 & 0 & 0 \\
 0 & 0 & 2 & 0 & 5 & 0 & 20 & 0 & 30 & 0 & 20 & 0 & 5 & 0 & 2 & 0 & 0 \\
 0 & 0 & 0 & 4 & 0 & 16 & 0 & 36 & 0 & 36 & 0 & 16 & 0 & 4 & 0 & 0 & 0 \\
 1 & 0 & 2 & 0 & 10 & 0 & 30 & 0 & 44 & 0 & 30 & 0 & 10 & 0 & 2 & 0 & 1 \\
 0 & 0 & 0 & 4 & 0 & 16 & 0 & 36 & 0 & 36 & 0 & 16 & 0 & 4 & 0 & 0 & 0 \\
 0 & 0 & 2 & 0 & 5 & 0 & 20 & 0 & 30 & 0 & 20 & 0 & 5 & 0 & 2 & 0 & 0 \\
 0 & 0 & 0 & 0 & 0 & 6 & 0 & 16 & 0 & 16 & 0 & 6 & 0 & 0 & 0 & 0 & 0 \\
 0 & 0 & 0 & 0 & 2 & 0 & 5 & 0 & 10 & 0 & 5 & 0 & 2 & 0 & 0 & 0 & 0 \\
 0 & 0 & 0 & 0 & 0 & 0 & 0 & 4 & 0 & 4 & 0 & 0 & 0 & 0 & 0 & 0 & 0 \\
 0 & 0 & 0 & 0 & 0 & 0 & 2 & 0 & 2 & 0 & 2 & 0 & 0 & 0 & 0 & 0 & 0 \\
 0 & 0 & 0 & 0 & 0 & 0 & 0 & 0 & 0 & 0 & 0 & 0 & 0 & 0 & 0 & 0 & 0 \\
 0 & 0 & 0 & 0 & 0 & 0 & 0 & 0 & 1 & 0 & 0 & 0 & 0 & 0 & 0 & 0 & 0
\end{array}
\right) \,,
\end{equation*}

\begin{equation*}
A_5 = \left(
\begin{array}{ccccccccccccccccccccc}
 0 & 0 & 0 & 0 & 0 & 0 & 0 & 0 & 0 & 0 & 1 & 0 & 0 & 0 & 0 & 0 & 0 & 0 & 0 & 0 & 0 \\
 0 & 0 & 0 & 0 & 0 & 0 & 0 & 0 & 0 & 0 & 0 & 0 & 0 & 0 & 0 & 0 & 0 & 0 & 0 & 0 & 0 \\
 0 & 0 & 0 & 0 & 0 & 0 & 0 & 0 & 2 & 0 & 2 & 0 & 2 & 0 & 0 & 0 & 0 & 0 & 0 & 0 & 0 \\
 0 & 0 & 0 & 0 & 0 & 0 & 0 & 0 & 0 & 4 & 0 & 4 & 0 & 0 & 0 & 0 & 0 & 0 & 0 & 0 & 0 \\
 0 & 0 & 0 & 0 & 0 & 0 & 2 & 0 & 5 & 0 & 10 & 0 & 5 & 0 & 2 & 0 & 0 & 0 & 0 & 0 & 0 \\
 0 & 0 & 0 & 0 & 0 & 0 & 0 & 6 & 0 & 18 & 0 & 18 & 0 & 6 & 0 & 0 & 0 & 0 & 0 & 0 & 0 \\
 0 & 0 & 0 & 0 & 2 & 0 & 6 & 0 & 24 & 0 & 39 & 0 & 24 & 0 & 6 & 0 & 2 & 0 & 0 & 0 & 0 \\
 0 & 0 & 0 & 0 & 0 & 6 & 0 & 28 & 0 & 60 & 0 & 60 & 0 & 28 & 0 & 6 & 0 & 0 & 0 & 0 & 0 \\
 0 & 0 & 2 & 0 & 5 & 0 & 24 & 0 & 68 & 0 & 101 & 0 & 68 & 0 & 24 & 0 & 5 & 0 & 2 & 0 & 0 \\
 0 & 0 & 0 & 4 & 0 & 18 & 0 & 60 & 0 & 120 & 0 & 120 & 0 & 60 & 0 & 18 & 0 & 4 & 0 & 0 & 0 \\
 1 & 0 & 2 & 0 & 10 & 0 & 39 & 0 & 101 & 0 & 144 & 0 & 101 & 0 & 39 & 0 & 10 & 0 & 2 & 0 & 1 \\
 0 & 0 & 0 & 4 & 0 & 18 & 0 & 60 & 0 & 120 & 0 & 120 & 0 & 60 & 0 & 18 & 0 & 4 & 0 & 0 & 0 \\
 0 & 0 & 2 & 0 & 5 & 0 & 24 & 0 & 68 & 0 & 101 & 0 & 68 & 0 & 24 & 0 & 5 & 0 & 2 & 0 & 0 \\
 0 & 0 & 0 & 0 & 0 & 6 & 0 & 28 & 0 & 60 & 0 & 60 & 0 & 28 & 0 & 6 & 0 & 0 & 0 & 0 & 0 \\
 0 & 0 & 0 & 0 & 2 & 0 & 6 & 0 & 24 & 0 & 39 & 0 & 24 & 0 & 6 & 0 & 2 & 0 & 0 & 0 & 0 \\
 0 & 0 & 0 & 0 & 0 & 0 & 0 & 6 & 0 & 18 & 0 & 18 & 0 & 6 & 0 & 0 & 0 & 0 & 0 & 0 & 0 \\
 0 & 0 & 0 & 0 & 0 & 0 & 2 & 0 & 5 & 0 & 10 & 0 & 5 & 0 & 2 & 0 & 0 & 0 & 0 & 0 & 0 \\
 0 & 0 & 0 & 0 & 0 & 0 & 0 & 0 & 0 & 4 & 0 & 4 & 0 & 0 & 0 & 0 & 0 & 0 & 0 & 0 & 0 \\
 0 & 0 & 0 & 0 & 0 & 0 & 0 & 0 & 2 & 0 & 2 & 0 & 2 & 0 & 0 & 0 & 0 & 0 & 0 & 0 & 0 \\
 0 & 0 & 0 & 0 & 0 & 0 & 0 & 0 & 0 & 0 & 0 & 0 & 0 & 0 & 0 & 0 & 0 & 0 & 0 & 0 & 0 \\
 0 & 0 & 0 & 0 & 0 & 0 & 0 & 0 & 0 & 0 & 1 & 0 & 0 & 0 & 0 & 0 & 0 & 0 & 0 & 0 & 0
\end{array}
\right)
\end{equation*}

\begin{equation*}
A_6 = \left(
\begin{array}{ccccccccccccccccccccccccc}
 0 & 0 & 0 & 0 & 0 & 0 & 0 & 0 & 0 & 0 & 0 & 0 & 1 & 0 & 0 & 0 & 0 & 0 & 0 & 0 & 0 & 0 & 0 & 0 & 0 \\
 0 & 0 & 0 & 0 & 0 & 0 & 0 & 0 & 0 & 0 & 0 & 0 & 0 & 0 & 0 & 0 & 0 & 0 & 0 & 0 & 0 & 0 & 0 & 0 & 0 \\
 0 & 0 & 0 & 0 & 0 & 0 & 0 & 0 & 0 & 0 & 2 & 0 & 2 & 0 & 2 & 0 & 0 & 0 & 0 & 0 & 0 & 0 & 0 & 0 & 0 \\
 0 & 0 & 0 & 0 & 0 & 0 & 0 & 0 & 0 & 0 & 0 & 4 & 0 & 4 & 0 & 0 & 0 & 0 & 0 & 0 & 0 & 0 & 0 & 0 & 0 \\
 0 & 0 & 0 & 0 & 0 & 0 & 0 & 0 & 2 & 0 & 5 & 0 & 10 & 0 & 5 & 0 & 2 & 0 & 0 & 0 & 0 & 0 & 0 & 0 & 0 \\
 0 & 0 & 0 & 0 & 0 & 0 & 0 & 0 & 0 & 6 & 0 & 18 & 0 & 18 & 0 & 6 & 0 & 0 & 0 & 0 & 0 & 0 & 0 & 0 & 0 \\
 0 & 0 & 0 & 0 & 0 & 0 & 2 & 0 & 6 & 0 & 25 & 0 & 41 & 0 & 25 & 0 & 6 & 0 & 2 & 0 & 0 & 0 & 0 & 0 & 0 \\
 0 & 0 & 0 & 0 & 0 & 0 & 0 & 6 & 0 & 30 & 0 & 68 & 0 & 68 & 0 & 30 & 0 & 6 & 0 & 0 & 0 & 0 & 0 & 0 & 0 \\
 0 & 0 & 0 & 0 & 2 & 0 & 6 & 0 & 30 & 0 & 92 & 0 & 134 & 0 & 92 & 0 & 30 & 0 & 6 & 0 & 2 & 0 & 0 & 0 & 0 \\
 0 & 0 & 0 & 0 & 0 & 6 & 0 & 30 & 0 & 98 & 0 & 200 & 0 & 200 & 0 & 98 & 0 & 30 & 0 & 6 & 0 & 0 & 0 & 0 & 0 \\
 0 & 0 & 2 & 0 & 5 & 0 & 25 & 0 & 92 & 0 & 230 & 0 & 317 & 0 & 230 & 0 & 92 & 0 & 25 & 0 & 5 & 0 & 2 & 0 & 0 \\
 0 & 0 & 0 & 4 & 0 & 18 & 0 & 68 & 0 & 200 & 0 & 368 & 0 & 368 & 0 & 200 & 0 & 68 & 0 & 18 & 0 & 4 & 0 & 0 & 0 \\
 1 & 0 & 2 & 0 & 10 & 0 & 41 & 0 & 134 & 0 & 317 & 0 & 429 & 0 & 317 & 0 & 134 & 0 & 41 & 0 & 10 & 0 & 2 & 0 & 1 \\
 0 & 0 & 0 & 4 & 0 & 18 & 0 & 68 & 0 & 200 & 0 & 368 & 0 & 368 & 0 & 200 & 0 & 68 & 0 & 18 & 0 & 4 & 0 & 0 & 0 \\
 0 & 0 & 2 & 0 & 5 & 0 & 25 & 0 & 92 & 0 & 230 & 0 & 317 & 0 & 230 & 0 & 92 & 0 & 25 & 0 & 5 & 0 & 2 & 0 & 0 \\
 0 & 0 & 0 & 0 & 0 & 6 & 0 & 30 & 0 & 98 & 0 & 200 & 0 & 200 & 0 & 98 & 0 & 30 & 0 & 6 & 0 & 0 & 0 & 0 & 0 \\
 0 & 0 & 0 & 0 & 2 & 0 & 6 & 0 & 30 & 0 & 92 & 0 & 134 & 0 & 92 & 0 & 30 & 0 & 6 & 0 & 2 & 0 & 0 & 0 & 0 \\
 0 & 0 & 0 & 0 & 0 & 0 & 0 & 6 & 0 & 30 & 0 & 68 & 0 & 68 & 0 & 30 & 0 & 6 & 0 & 0 & 0 & 0 & 0 & 0 & 0 \\
 0 & 0 & 0 & 0 & 0 & 0 & 2 & 0 & 6 & 0 & 25 & 0 & 41 & 0 & 25 & 0 & 6 & 0 & 2 & 0 & 0 & 0 & 0 & 0 & 0 \\
 0 & 0 & 0 & 0 & 0 & 0 & 0 & 0 & 0 & 6 & 0 & 18 & 0 & 18 & 0 & 6 & 0 & 0 & 0 & 0 & 0 & 0 & 0 & 0 & 0 \\
 0 & 0 & 0 & 0 & 0 & 0 & 0 & 0 & 2 & 0 & 5 & 0 & 10 & 0 & 5 & 0 & 2 & 0 & 0 & 0 & 0 & 0 & 0 & 0 & 0 \\
 0 & 0 & 0 & 0 & 0 & 0 & 0 & 0 & 0 & 0 & 0 & 4 & 0 & 4 & 0 & 0 & 0 & 0 & 0 & 0 & 0 & 0 & 0 & 0 & 0 \\
 0 & 0 & 0 & 0 & 0 & 0 & 0 & 0 & 0 & 0 & 2 & 0 & 2 & 0 & 2 & 0 & 0 & 0 & 0 & 0 & 0 & 0 & 0 & 0 & 0 \\
 0 & 0 & 0 & 0 & 0 & 0 & 0 & 0 & 0 & 0 & 0 & 0 & 0 & 0 & 0 & 0 & 0 & 0 & 0 & 0 & 0 & 0 & 0 & 0 & 0 \\
 0 & 0 & 0 & 0 & 0 & 0 & 0 & 0 & 0 & 0 & 0 & 0 & 1 & 0 & 0 & 0 & 0 & 0 & 0 & 0 & 0 & 0 & 0 & 0 & 0
\end{array}
\right)
\end{equation*}

\bibliography{bib_alg}

\end{document}